\documentclass[reprint,onecolumn,amsmath,amssymb,aps,nofootinbib,pre]{revtex4-2}
\usepackage{graphicx,hyperref}
\usepackage{bm}
\usepackage{comment}
\usepackage{color}
\allowdisplaybreaks

\begin{document}
\title{Mpemba effect in a two-dimensional bistable potential}
\author{Hisao Hayakawa}
\email[e-mail: ]{hisao@yukawa.kyoto-u.ac.jp}
\affiliation{Center for Gravitational Physics and Quantum Information, Yukawa Institute for Theoretical Physics, Kyoto University, Kitashirakawa-Oiwakecho, Sakyo-ku, Kyoto 606-8502, Japan}
\author{Satoshi Takada}
\email[e-mail: ]{takada@go.tuat.ac.jp}
\affiliation{Department of Mechanical Systems Engineering and 
    Institute of Engineering, 
    Tokyo University of Agriculture and Technology, 
    2-24-16 Naka-cho, Koganei, Tokyo 184-8588, Japan}
\date{\today}
\begin{abstract}
We present an exactly solvable model of the Mpemba effect in an overdamped Langevin system confined to a two-dimensional, radially symmetric bistable potential. 
The potential is constructed as a piecewise quadratic-logarithmic function that is continuous and differentiable at the matching radii, enabling an exact mapping of the corresponding Fokker-Planck operator to a Schr\"{o}dinger-type eigenvalue problem. 
The relaxation spectrum and eigenmodes are obtained analytically in each region in terms of confluent hypergeometric functions, with eigenvalues determined from matching conditions.
Focusing on isotropic equilibrium initial states at inverse temperature $\beta_{\rm ini}$ quenched to a bath at inverse temperature $\beta$, we derive explicit expressions for the mode amplitudes governing long-time relaxation. 
We demonstrate that the coefficient of the slowest mode exhibits a non-monotonic dependence on $\beta_{\rm ini}$ and identify a sufficient crossing condition for the Kullback-Leibler divergence in terms of the two slowest modes, if the global minimum of the potential is located far away from the origin and the second minimum exists near the origin. 
For corresponding parameters, we demonstrate that the Mpemba effect can be realized.
Our results provide a rare example of an analytically tractable two-dimensional model exhibiting anomalous relaxation without any confining walls, extending previous one-dimensional constructions with a hard wall and clarifying the role of radial geometry in nonequilibrium relaxation phenomena.
\end{abstract}
\maketitle

\section{Introduction}
The Mpemba effect refers to the anomalous relaxation phenomenon in which an initially hotter system,
quenched into contact with a cold bath, reaches equilibrium faster than the same system prepared at a
lower initial temperature. The effect was first reported by Mpemba and Osborne \cite{Mpemba1969}, who
observed that hotter water can freeze faster than colder water. Although its very existence has been
debated \cite{Burridge2016,Katz2017}, theoretical studies soon after the debates already suggested
possible scenarios for faster freezing from higher temperatures \cite{Lu2017,Lasanta2017}. In recent
years, numerous experimental and theoretical works have confirmed Mpemba-like processes in which a
``temperature-like'' observable exhibits a crossing, such that a system initialized at higher
temperature subsequently relaxes faster than the same system started at lower temperature.

The Mpemba effect has been reported in a broad spectrum of classical systems, including colloidal
particles in optical traps \cite{Kumar2020,Kumar2022,Malhotra24}, 
nanoparticles in optical traps \cite{Tian2025},
thermostated granular fluids \cite{Lasanta2017,Torrente2019,Biswas2020,Biswas2021,Mompo2021,Megias2022,Patron2023,Biswas23}, optical
resonators \cite{Keller2018,Santos2020,Patron2021}, inertial suspensions \cite{Takada2021,Takada2021b},
spin glasses \cite{Baity2019}, molecular binary mixtures \cite{Gonzalez2021}, phase transitions~\cite{Holtzman2022,S_Chatterjee24,Li2026}, and a wide
variety of Markovian model
\cite{Lu2017,Santos2024,Klich2019,Busiello2021,Lin2022,Pemartin23}, Langevin model or equivalent Fokker-Planck model \cite{Biswas2023,Biswas_thesis,Deguenther22,Liu26,Liu26long}, non-Markovian models~\cite{Strachan2025,Li2025} and active matter~\cite{Biswas2025,Antonov26}. In
addition, quantum analogues of the effect have been demonstrated in open quantum systems
\cite{Nava2019,Carollo2021,Manikandan2021,Ivander2023,Ares2023,Chatterjee2023,Chatterjee2024,Joshi2024,
Shapira2024,Rylands2024,Moroder2024,Chalas2024,Ares2025,Yamashika2024,Liu2024,Wang2024,Nava2024,
Longhi2024,Turkeshi2025,Bao2025,Yamashika2026,Yamashika26,Hayakawa2026,Melles26}, establishing the Mpemba effect as a general nonequilibrium relaxation phenomenon.

Several reviews now provide unified perspectives on both classical and quantum Mpemba effects
\cite{Bechhoefer2021,Teza2025,Ares2025Rev,Yu2025}, emphasizing their connection to nonequilibrium speed limits
and anomalous relaxation mechanisms. 
See also a recent paper for a unified view on the Mpemba effect~\cite{Summer26}.
Two broad scenarios are typically distinguished
\cite{Liu26,Liu26long,Ohga2024,VVu2025}: (i) comparing relaxation from an equilibrium versus a
non-equilibrium initial condition, and (ii) comparing relaxation from two equilibrium states prepared
at different initial temperatures. The latter scenario, which we follow in this paper, has led to
general unifying frameworks \cite{Ohga2024,VVu2025}.

A central theoretical picture, proposed by Lu and Raz \cite{Lu2017}, is that the Mpemba effect arises in systems with a non-convex free-energy landscape. Starting from an initial distribution localized near a metastable minimum, the mean relaxation time across the barrier decreases with increasing initial temperature, leading to the observed crossing phenomenon. This interpretation naturally
extends to one-particle dynamics in bistable potentials, as confirmed in experiments on optically trapped colloids \cite{Kumar2020,Kumar2022}. 
Variants such as the strong Mpemba effect \cite{Klich2019} have also been studied, though their relation to the original freezing anomaly is more tenuous.
If we are interested in relaxations from equilibrium initial conditions, the Mpemba effect arises when the projection onto the slowest mode is smaller for the initially hotter state, even if its total distance from equilibrium is larger.

So far, most studies discussed the motion of a particle in a discrete-level model, piecewise linear potentials, and square box potentials~\cite{Lu2017,Lasanta2017,Kumar2020,Walker2021,Biswas2023,Biswas23s}.
For instance, Lu and Raz~\cite{Lu2017} investigated the effect using discrete states and numerical analyses of the Master equation.
Kumar and Bechhofer~\cite{Kumar2020} provided an important experimental realization complemented by numerical calculations of the probability density.
Walker and Vucelja~\cite{Walker22} employed a semi-analytical treatment, but it is fundamentally based on the high-barrier Kramers limit and supplemented by numerical calculations.
None of these approaches provides the exact, explicit spectrum and mode amplitudes derived in our manuscript.
Quite recently, Liu et al.~\cite{Liu26,Liu26long} developed a systematic theory of the Mpemba effect of particles confined in one-dimensional single-well and double-well potentials, but they also did not write an explicit solution except for the symmetric double-well potential case.

Experimentally and theoretically, one-dimensional systems differ fundamentally from higher-dimensional systems, such as two-dimensional and three-dimensional systems.
To ground our theoretical analysis in concrete physical contexts, we emphasize that two-dimensional radially symmetric bistable potentials naturally emerge across several prominent experimental platforms. 
In soft matter and colloidal physics, such landscapes are routinely microengineered using advanced optical tweezers equipped with spatial light modulators or rapidly scanning acousto-optic deflectors. 
These instruments can sculpt a central localized trapping well surrounded by a concentric, ring-shaped optical barrier or secondary outer well, providing an ideal, highly controllable testbed for studying two-dimensional Brownian motion. Radial bistability also arises as an effective free-energy landscape in structural phase transitions, such as the radial packing of lipid domains in biomembranes or topological defect relaxation in liquid crystals confined to cylindrical geometries. 
On nanoscale dimensions, a similar potential energy profile describes the conduction band offset in core-shell quantum disks or concentric semiconductor heterostructures. By analyzing the Mpemba effect in these specific two-dimensional radially symmetric setups, our work offers explicit kinetic criteria that can be directly leveraged to optimize or anomalously accelerate relaxation rates in micro-to-nanoscopic devices via strategic initial thermal quenches.

Recently, several papers~\cite{Walker2021,Biswas23s,Liu26,Liu26long} found that the metastable picture is not correct for realizing the Mpemba effect. 
Still, a confining wall is necessary to observe the Mpemba effect. 
To understand the robustness of the Mpemba effect, it is natural to analyze a two-dimensional system to clarify the differences and common features in one-dimensional and two-dimensional systems, including whether confining walls are necessary to observe the Mpemba effect even in two dimensions.

In this work, we develop an exactly solvable model of the Mpemba effect in a \emph{smooth two-dimensional asymmetric bistable potential} by using a connection of three potentials, taking into account matching conditions at the connection points. 
While previous exact or semi-analytical approaches have mostly been restricted to one-dimensional models, our extension to two dimensions provides a natural and richer generalization. 
The analysis makes use of the mapping from the Fokker-Planck equation to a Schr\"odinger-type eigenvalue problem, which enables us to determine the relaxation spectrum and eigenmodes exactly. 
We show that the Mpemba effect of the Kullback-Leibler (KL) divergence~\cite{sagawa}, which is the best monotone measure, i.e., decreases monotonically with time (and its inverse) can be realized in this setting without the introduction of any confining walls, thus broadening the landscape of analytically tractable nonequilibrium phenomena.
Unlike previous criteria based on specific observables, our condition is formulated directly in terms of the full probability distribution via the KL divergence.

The contents of this paper are as follows.
In the next section, we explain the model and setup, including the introduction of a solvable potential and the mapping onto the Schr\"{o}dinger equation.
In Sec.~\ref{Sec:Mode-analysis}, we present the mode analysis of the mapped Schr\"{o}dinger equation to obtain the solution of Schr\"{o}dinger equation.
In Sec.~\ref{Sec:Demo-Mpemba}, we demonstrate the occurrence of the Mpemba effect by using the two-mode approximation.
In Sec.~\ref{Sec:Discussion}, we discuss our results.
In Sec.~\ref{Sec:Conclusion}, we present a summary of our results.
In Appendix~\ref{app:potential}, we present the detailed form of the potential used for the analysis.
In Appendix~\ref{app:hypergeometric}, we briefly summarize the properties of hypergeometric functions we use.
In Appendix~\ref{app:Schroedinger}, we present the detailed solution of the Schr\"{o}dinger
equation, including the matching conditions and the determination of
the eigenmodes.
The detailed analysis of the marginal case $V(\alpha)=0$ is presented in Appendix~\ref{marginal}.
In Appendix~\ref{App:crossing}, we present the condition for the crossing of the KL divergence.
In Appendix~\ref{App:Higher}, we discuss the contribution of higher-order terms of mode analysis.
In Appendix~\ref{app:system_size}, we illustrate that $a_2(\beta_\mathrm{ini})$ is independent of the system size.
In Appendix~\ref{app:F_2}, we present the detailed conditions for having a peak of the slowest eigenmode against the initial temperature, mathematically. 
The numerical codes used in this study are publicly available in Ref.~\cite{code}.

\section{Model and setup}\label{Sec:model}
\subsection{Formulation}

\begin{figure}[htbp]
    \centering
    \includegraphics[width=0.4\linewidth]{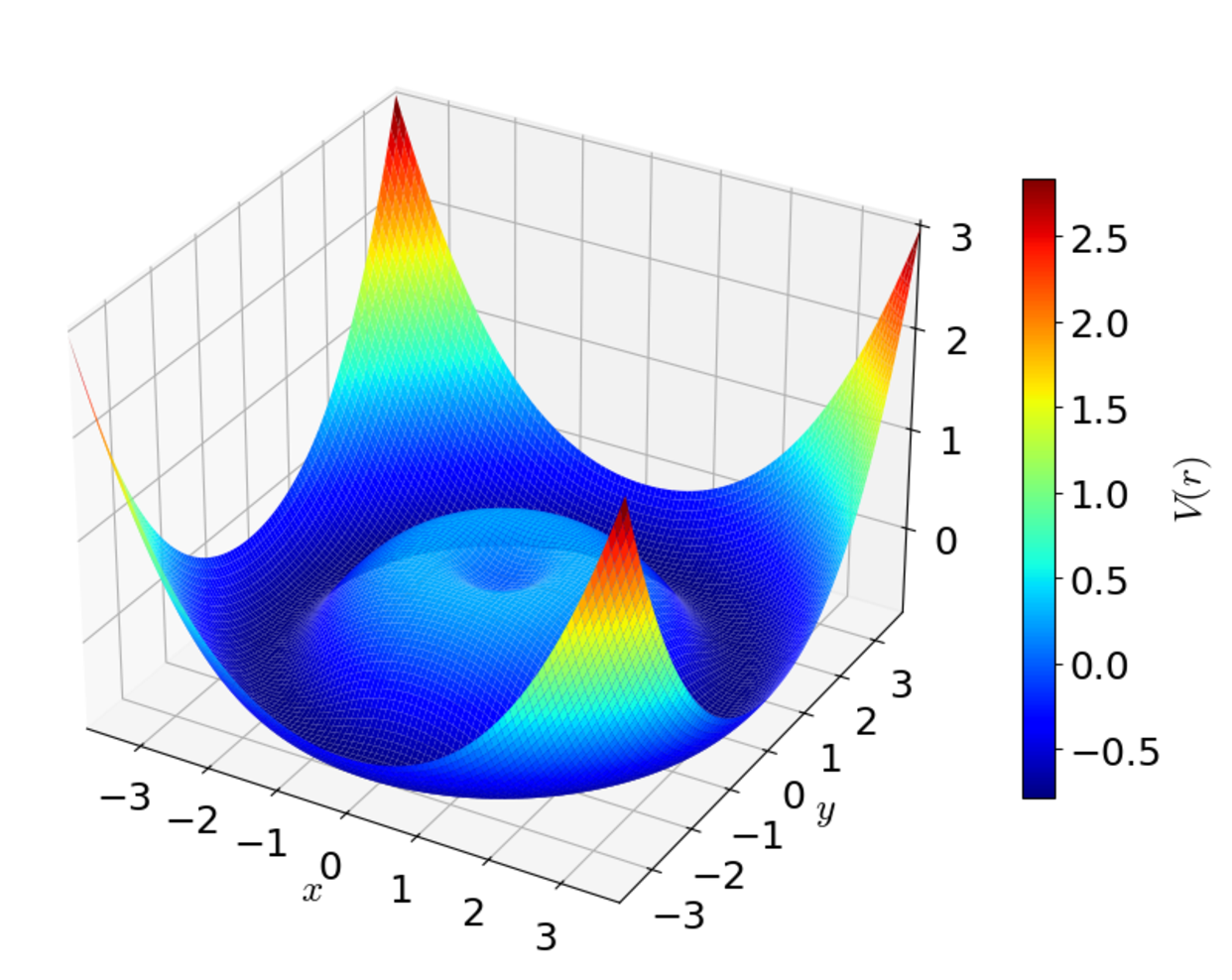}
    \caption{A schematic of a bistable radially symmetric potential for $k_\mathrm{in}=1$, $k_\mathrm{mid}=-0.5$,  $k_\mathrm{out}=0.8$, $\xi=1$, and $\alpha=3$.}
    \label{fig:pot}
\end{figure}

Let us consider the motion of a particle at position $\bm{r}$ in a two-dimensional system.
Here, the particle moves under the influence of a bistable potential $V(r)$ with $r:=|\bm{r}|$ (see Fig.~\ref{fig:pot}).
We assume that the motion of the particle obeys the dimensionless Langevin equation as
\begin{equation}
    \dot{\bm{r}}
    =- \frac{\partial V(r)}{\partial \bm{r}}
    + \bm{\xi}(t) ,
    \label{eq:Langevin_cross}
\end{equation}
where $\dot{\bm{r}}:=d\bm{r}(t)/dt$, and the noise $\xi_\alpha$, which is $\alpha$ component of $\bm{\xi}$, satisfies
\begin{equation}
    \langle \xi_\alpha(t)\rangle = 0,\quad
    \langle \xi_\alpha(t) \xi_\beta(t^\prime)\rangle
    = 2 T
    \delta_{\alpha\beta}\delta(t-t^\prime)
\end{equation}
with $T$ being the bath temperature.

The corresponding Fokker-Planck equation can be written as
\begin{align}\label{m-Smoluchowski2}
    \frac{\partial P(\bm{r},t)}{\partial t}&=\mathbb{L}P(\bm{r},t) , \\
    \mathbb{L}:&=\bm{\nabla}\cdot \left( \frac{\bm{r}}{r}V'(r)\right)+T \nabla^2,
    \label{def:L}
\end{align}
where $V'(r):=dV(r)/dr$.
The analysis of the Fokker-Planck equation follows the textbook~\cite{risken1996fokker}.

Equations \eqref{m-Smoluchowski2} and \eqref{def:L} can be rewritten as
\begin{align}\label{current}
     \frac{\partial P(r,t)}{\partial t}&=-\bm{\nabla}\cdot \bm{J}(\bm{r},t), \quad
     \bm{J}(\bm{r},t):=-\left\{T\bm{\nabla} +\frac{\bm{r}}{r}V'(r)\right\}P(\bm{r},t).
\end{align}
Note that $P(\bm{r},t)$ should be isotropic if the initial condition is isotropic, i.e. $P(\bm{r},t)=P(r,t)$ because the dynamics is governed by a radially symmetric potential $V(r)$.

We consider a quench process from $T_\mathrm{ini}=\beta_\mathrm{ini}^{-1}$ to $T=\beta^{-1}$.
We also assume that the initial and final states are at equilibrium as 
\begin{align}\label{P_{ini}}
    P_\mathrm{ini}(r)&=P_\mathrm{eq}(r, \beta_\mathrm{ini}):=\frac{e^{-\beta_\mathrm{ini}V(r)}}{Z(\beta_\mathrm{ini})},\\
    P(r,t\to \infty)&=P_\mathrm{eq}(r,\beta),
\end{align}
with $Z(\beta):=2\pi \int_0^\infty dr r e^{-\beta V(r)}$, and $P_\mathrm{eq}(r,\beta)$ satisfies
\begin{align}
    \mathbb{L} P_\mathrm{eq}(r,\beta)=0 .
\end{align}

\subsection{A choice of solvable potential}

In general, it is difficult to obtain the exact solution of the Fokker-Planck equation with a double-well potential.
However, for a suitably chosen form of the potential $V(r)$, we can solve the Fokker-Planck equation.
We therefore construct an exactly solvable radially symmetric double-well potential $V(r)$ in two dimensions. 
The requirements are:
1) $V(r)$ is continuous and differentiable across its domains,
2) $V(r)$ has a minimum at $r=0$, a maximum at $r=\xi$, and a second minimum at $r=\alpha>\xi$.

We take $V(r)$ the piecewise quadratic-logarithmic form:
\begin{align}\label{pot2}
    V(r)=
    \begin{cases}
    V_{\rm in}(r)=\dfrac{k_{\rm in}}{2} r^2 + C_{\rm in}, & 0\le r < r_- , \\[0.5em]
    V_{\rm mid}(r)=\dfrac{k_{\rm mid}}{2} r^2 + b_{\rm mid}\ln r + C_{\rm mid}, & r_- \le r < r_+, \\[0.5em]
    V_{\rm out}(r)=\dfrac{k_{\rm out}}{2} r^2 + b_{\rm out}\ln r + C_{\rm out}, & r \ge r_+ ,
    \end{cases}
\end{align}
where $r_\pm$ denote matching points, and $C_{\rm in},C_{\rm mid}$, and $C_{\rm out}$ are, respectively, constants to ensure continuity.
As will be shown, the conditions $b_\mathrm{mid}\propto \xi^2$ and $b_\mathrm{out}\propto \alpha^2$ should be satisfied.
Here, the explicit expressions of $r_-$, $r_+$, $C_\mathrm{mid}$, and $C_\mathrm{out}$ are presented in Appendix \ref{app:potential}.
Notably, this specific form leads to an exactly solvable Fokker-Planck equation.
We also summarize the parameters appearing in the model in Table \ref{tab:parameters}.
Here, we change the values of $k_\mathrm{in}$, $k_\mathrm{mid}$, $k_\mathrm{out}$, $\xi$, and $\alpha$ depending on the cases and are referred as \textit{Model parameters}.
Once we fix these parameters, $r_-$ and $r_+$ are determined by Eq.~\eqref{eq:r_-_r_+} and $C_\mathrm{mid}$ and $C_\mathrm{out}$ by Eqs.~\eqref{C_mid} and \eqref{C_out}, respectively.

\begin{table}[htbp]
    \caption{List of the variables used in this paper.
    The meaning of \textit{Model parameter} is explained in the main text.}
    \label{tab:parameters}
    \begin{ruledtabular}
    \begin{tabular}{lll}
    Symbol & Meaning & Determination \\ \hline
    $T$ & Bath temperature & Control parameter \\
    $\beta=1/T$ & Inverse bath temperature & Defined from $T$ \\
    $\beta_{\rm ini}$ & Initial inverse temperature & Initial condition \\
    $k_{\rm in}$ & Curvature of the inner well & Model parameter$^\dagger$ \\
    $k_{\rm mid}$ & Curvature of the barrier region & Model parameter$^\dagger$ \\
    $k_{\rm out}$ & Curvature of the outer well & Model parameter$^\dagger$ \\
    $\xi$ & Position of the barrier top & Model parameter$^\dagger$ \\
    $\alpha$ & Position of the outer minimum & Model parameter$^\dagger$ \\
    $b_{\rm mid}$ & Logarithmic coefficient in $V_{\rm mid}$ & $-k_{\rm mid}\xi^2$ \\
    $b_{\rm out}$ & Logarithmic coefficient in $V_{\rm out}$ & $-k_{\rm out}\alpha^2$ \\
    $r_{-}$ & Inner matching radius & Eq.~\eqref{eq:r_-_r_+} \\
    $r_{+}$ & Outer matching radius & Eq.~\eqref{eq:r_-_r_+} \\
    $C_{\rm in}$ & Constant shift of $V_{\rm in}$ & Arbitrary (set to $0$) \\
    $C_{\rm mid}$ & Constant ensuring continuity & Eq.~\eqref{C_mid} \\
    $C_{\rm out}$ & Constant ensuring continuity & Eq.~\eqref{C_out}
    \end{tabular}
    \end{ruledtabular}
\end{table}

We distinguish three cases: (i) $V(\alpha)<0$, (ii) $V(\alpha)>0$, and (iii) $V(\alpha)=0$ for later discussion.
For the representative choice $k_\mathrm{in}=k_\mathrm{out}=\xi=1$, the marginal line for $V(\alpha)=0$ is shown in Fig.~\ref{fig:marginal}.
See Appendix \ref{app:potential} for the detailed form of the potential analyzed in this paper, including the choice of the parameters.

\begin{figure}[htbp]
    \centering
    \includegraphics[width=0.5\linewidth]{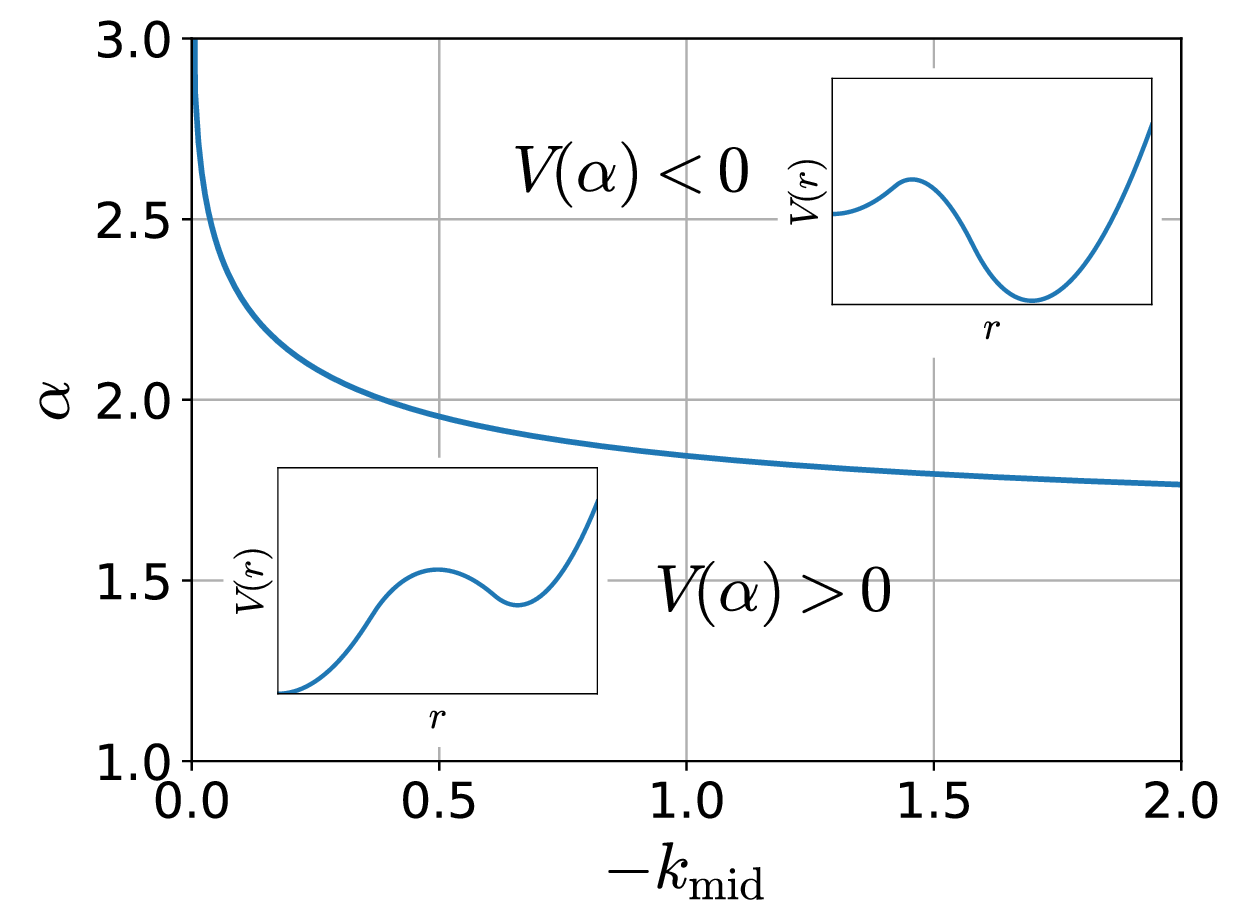}
    \caption{Phase diagram of the sign of $V(\alpha)$ in the parameter space $(-k_\mathrm{mid},\alpha)$ for $k_\mathrm{in}=k_\mathrm{out}=\xi=1$, where the solid line represents the boundary defined by $V(\alpha)=0$.
    The insets show schematic pictures of the potential corresponding to both cases ($(-k_\mathrm{mid},\alpha)=(0.5, 1.5)$ for $V(\alpha)>0$ and $(1.5, 2.5)$ for $V(\alpha)<0$, respectively).
    }
    \label{fig:marginal}
\end{figure}

\subsection{Map to Schr\"{o}dinger equation}
\subsubsection{Framework}
We can map from the Fokker-Planck equation to Schr\"{o}dinger equation.
This is because the Fokker-Planck operator $\mathbb{L}$ is a non-Hermitian operator, while a Hermitian operator describes the Schr\"{o}dinger equation.
Thus, this mapping helps us to get concrete results in terms of well-known properties of the Hermitian operators. 

Let us introduce $\mathbb{P}(\bm{r},t)$ defined as 
\begin{align}\label{map}
    \mathbb{P}(r,t):=e^{\beta V(r)/2}P(r,t) ,
\end{align}
where $\mathbb{P}(\bm{r},t)$ satisfies
\begin{align}\label{S-Eq}
    \frac{\partial}{\partial t}\mathbb{P}(r,t)&=-\mathbb{H}\mathbb{P}(r,t) ,\\
    \mathbb{H}&:=-e^{\beta V(r)/2}\mathbb{L}e^{-\beta V(r)/2}=
    -T\nabla^2+V_S(r).
\label{mathbb_L}
\end{align}

Recall that the Schr\"{o}dinger mapping gives
\begin{align}
    V_S(r) := \frac{1}{4T}\, 
    (V'(r))^2 - \frac{1}{2r}\frac{d}{dr}\!\big(r V'(r)\big).
\end{align}
For the potential in Eq.~\eqref{pot2}, $V_S(r)$ is given by
\begin{align}
    V_S(r) = 
    \begin{cases}
    -k_{\rm in} +  \dfrac{k_{\rm in}^2}{4T} r^2 , & 0\le r< r_- , \\[1.0em]
    -k_{\rm mid} + \dfrac{k_{\rm mid} b_{\rm mid}}{2T} + \dfrac{k_{\rm mid}^2}{4T} r^2 + \dfrac{b_{\rm mid}^2}{4T r^2}, & r_- \le r < r_+ , \\[1.0em]
    -k_{\rm out} + \dfrac{k_{\rm out} b_{\rm out}}{2T} + \dfrac{k_{\rm out}^2}{4T} r^2 + \dfrac{b_{\rm out}^2}{4T r^2}, & r \ge r_+ .
    \end{cases}
    \label{V_S}
\end{align}
Requiring that $V_S(r)$ has extrema at $r=0$, $\xi$, and $\alpha$ yields
\begin{align}   
    b_{\rm mid}=-k_{\rm mid}\,\xi^2, \quad
    b_{\rm out}=-k_{\rm out}\,\alpha^2.
    \label{eq:bin_bmid_bout}
\end{align}
Equation~\eqref{V_S} can be rewritten as
\begin{align}
    V_S(r)= -k_\mathrm{I}+\frac{k_\mathrm{I} b_\mathrm{I}}{2T}
    +\frac{k_\mathrm{I}^2}{4T}r^2+\frac{b_\mathrm{I}^2}{4T r^2} ,
\end{align}
where the suffix I represents the region $\mathrm{I}\in\{\mathrm{in},\mathrm{mid},\mathrm{out}\}$ with $b_\mathrm{in}=0$.
The advantage of using the map to Schr\"{o}dinger equation is obvious because of $\mathbb{H}^\dagger=\mathbb{H}$.

We note the following:
i) Each segment of $V_S(r)$ is solvable quadratic plus inverse-square potential. 
ii) The solutions of the corresponding radial Schr\"{o}dinger equation are expressible in terms of confluent hypergeometric functions. 
iii) The matching conditions at $r=r_-$ and $r=r_+$ enforce continuity of both $V(r)$ and $V'(r)$, which automatically ensure the continuity of $V_S(r)$ and the physical wavefunctions.
iv) The smooth matching conditions for $\mathbb{P}(r,t)$ and $\partial_r\mathbb{P}(r,t)$ at $r=r_-$ and $r=r_+$ provide a globally smooth solution of Eq. \eqref{S-Eq} because the analyzed equation becomes a second-order ordinary differential equation.
This construction provides a fully solvable model for a two-dimensional bistable potential with spherical symmetry.

Now, the problem is mapped onto the eigenvalue problem of the operator $\mathbb{H}$: 
\begin{align}\label{Eq32}
    \mathbb{H}\varphi_{m,n}(r)
    =\lambda_{m,0} \varphi_{m,n}(r) ,
\end{align}
where $\varphi_{m,n}(r)$ is related to the left and right eigenfunctions as
\begin{equation}\label{lrphi}
    \ell_{m,n}(\bm{r})
    =e^{\beta V(r)/2}\varphi_{m,n}(r), \quad \mathrm{and}\quad r_{m,n}(r)
    =e^{-\beta V(r)/2}\varphi_{m,n}(r) ,
\end{equation}
where $\ell_{m,n}(\bm{r})$ and $r_{m,n}(\bm{r})$ are, respectively, the left and right eigenvectors, which satisfy
\begin{align}\label{left_eigen}
    \mathbb{L}r_{m,n}(\bm{r})&=-\lambda_{m,0} r_{m,n}(\bm{r}) ,\\
    \mathbb{L}^\dagger \ell_{m,n}(\bm{r})&=-\lambda_{m,0} \ell_{m,n}(\bm{r}), \quad  \mathbb{L}^\dagger:=-\frac{\bm{r}}{r}V'(r)\cdot\bm{\nabla}+T \nabla^2 .
\label{right_eigen}
\end{align}
Note that the two-dimensional Schr\"{o}dinger equation has discrete eigenvalues characterized by a set of non-negative integers $(m,n)$, where the eigenvalue for an isotropic potential $V_S(r)$ should be independent of $n$, i.e., $\lambda_{m,n}=\lambda_{m,0}$ for an arbitrary non-negative integer $n$.

When we begin with an isotropic (equilibrium) condition Eq.~\eqref{P_{ini}}, the solution of Schr\"{o}dinger equation is independent of the second index $n$.
Thus, the orthonormal relation 
\begin{align}\label{orthnormal}
    \int d\bm{r}\ell_{m,n}(r)r_{m',n'}(r)
    =\delta_{mm'}\delta_{nn'}   
\end{align}
is converted into
\begin{align}\label{orthonormal2}
    \int d\bm{r} \varphi_{m,0}(r)\varphi_{m',0}(r)
    =\delta_{mm'}.
\end{align}
Then, we can use the spectrum decomposition of $P(r,t)$ as
\begin{align}\label{P(r,t)2}
    P(r,t)
    &= P_\mathrm{eq}(r,\beta)
    + e^{-\beta V(r)/2} 
    \sum_{m=2}^\infty a_{m}\phi_{m} 
    e^{-\tilde{\lambda}_{m}t} \notag\\
    &\approx P_\mathrm{eq}(r,\beta)
    +e^{-\beta V(r)/2}\left\{a_{2} \phi_{2}(r) e^{-\tilde{\lambda}_{2}t}
    +a_{3} \phi_{3}(r) e^{-\tilde{\lambda}_{3}t}
    \right\},
\end{align}
where
\begin{align}
       a_{m}&=\frac{2\pi\delta_{n0}}{Z(\beta_\mathrm{ini})}\int_0^\infty dr r \exp\left[\left(\frac{\beta}{2}-\beta_\mathrm{ini}\right)V(r)\right]\phi_{m}(r).
    \label{a_{mn}}
\end{align}
For such an initial condition, we shall introduce
\begin{align}
    \phi_{m}(r):=\varphi_{m-1,0}(r), \quad  \tilde{\lambda}_m:=\lambda_{m-1,0}=\lambda_{m-1,n} .
\end{align}
Note that $a_m$ corresponds to a one-dimensional counterpart, where $m=1$ and $m=2$ represent the zero mode and slowest relaxation mode, respectively. 
We also write the equation:
\begin{align}\label{F(beta)}
    F_m(\beta_\mathrm{ini}):=\frac{\partial a_m}{\partial \beta_\mathrm{ini}}= 
    -2\pi \int_0^\infty dr r e^{\beta V(r)/2}\phi_m(r)[V(r)-\langle V\rangle_{\beta_\mathrm{ini}}]P_\mathrm{ini}(r) 
\end{align}
with $\langle V\rangle_{\beta_\mathrm{ini}}:=2\pi\int_0^\infty dr r V(r) P_\mathrm{ini}(r)$, 
which is important to detect the necessary condition $F_2(\beta_\mathrm{ini}^*)=0$ at $\beta_\mathrm{ini}=\beta_\mathrm{ini}^*$. 

The approximate expression in Eq.~\eqref{P(r,t)2} is useful in the long-time limit.
Although the eigenvalues are degenerate for $n\ne 0$, the eigenvalues are non-degenerate for modes with $n=0$.
Thus, the eigenvalues satisfy the order $\tilde{\lambda}_1=0<\tilde{\lambda}_2<\tilde{\lambda}_3<\tilde{\lambda}_4<\cdots$ with finite gaps between $\tilde{\lambda}_n$ and $\tilde{\lambda}_{n+1}$. 
Therefore, the late-stage dynamics is dominated by the slowest modes associated with $\tilde{\lambda}_2$ and $\tilde{\lambda}_3$,

Using Eq.~\eqref{P(r,t)2}, we can write that the late-stage relaxation dynamics of the expectation value of an arbitrary observable $\hat{A}$ is described by
\begin{align}\label{<A>}
    \langle \hat{A} \rangle(t)\approx 2\pi\int_0^\infty dr r P_\mathrm{eq}(r,\beta)\hat{A}+2\pi \int_0^\infty dr r \hat{A}\left[e^{-\beta V(r)/2}\left\{a_{2} \phi_{2}(r) e^{-\tilde{\lambda}_{2}t}
    +a_{3} \phi_{3}(r) e^{-\tilde{\lambda}_{3}t}\right\}\right] ,
\end{align}
where $\langle \hat{A}\rangle(t):=2\pi\int_0^\infty dr r P(r,t)\hat{A}$.
Thus, the relaxation of $\langle \hat{A} \rangle(t)$ is dominated by the slow modes, proportional to $a_2$ and $a_3$.

Note that a one-dimensional model can be solved using a set of harmonic and anti-harmonic potentials~\cite{Liu26long}.
The idea of using harmonic and anti-harmonic potentials to describe tunneling in an asymmetric double-well potential in quantum mechanics was introduced by Dekker~\cite{Dekker87} and later developed by Song~\cite{Song08,Song15}.
Our analysis shares similarities with previous papers~\cite{Dekker87,Song08,Song15}, the target is completely different from them.

\subsubsection{Condition to observe the Mpemba effect}

Let us consider Eqs.~\eqref{P(r,t)2} and \eqref{<A>} in the large $t$ regime, where the relaxation term proportional to $a_2$ is only relevant.
Then, the previous papers~\cite{Lu2017,Teza2025,Liu26,Liu26long} have shown that the condition to observe the Mpemba effect is that $a_2$ has a maximum or a minimum with respect to $\beta_\mathrm{ini}$.
Indeed, this condition is necessary because the peak of $a_2$ at $\beta_\mathrm{ini}^*$ corresponds to the slowest initial temperature to approach the final equilibrium state after the quench.
This suggests that two relaxation processes, one starting from $\beta_\mathrm{ini}^*$ and the other from $\beta_\mathrm{ini}<\beta_\mathrm{ini}^*$, may cause a crossing of a thermodynamic observable.
However, this condition may be insufficient, as it does not guarantee that the slowest eigenmode crosses another relaxation mode proportional to $a_3$.
To improve the defect of the previous condition, which only $a_2$, we will propose a more direct crossing condition for a monotonic measure, such as the KL divergence, to observe the Mpemba effect later.

\subsubsection{Matching conditions}

Since we adopt the connected potential Eq.~\eqref{pot2}, which can be converted into the effective potential Eq.~\eqref{V_S}, we should impose the connection conditions at $r=r_-$ and $r=r_+$.
The connection conditions are given by
\begin{align}
    P_\mathrm{in}(r_-;t)&=P_\mathrm{mid}(r_-;t), \quad J_{r,\mathrm{in}}(r_-;t)=J_{r,\mathrm{mid}}(r_-;t) \\
    P_\mathrm{mid}\left(r_+;t\right)&=P_\mathrm{out}\left(r_+;t\right),
    \quad 
    J_{r,\mathrm{mid}}\left(r_+;t\right)=J_{r,\mathrm{out}}\left(r_+;t\right) ,
\end{align}
where $P_\mathrm{in}(r;t)$, $P_\mathrm{mid}(r;t)$, and $P_\mathrm{out}(r;t)$ are $P(r,t)$ for $0\le r<r_-$, $r_-\le r<r_+$, and $r\ge r_+$, respectively.
Similarly, $J_r(r,\theta;t):=\bm{e}_r \cdot \bm{J}(r,\theta;t)$ is the radial component of the current $\bm{J}$ introduced in Eq.~\eqref{current}, and
$J_{r,\mathrm{in}}$, $J_{r,\mathrm{mid}}$, and $J_{r,\mathrm{out}}$ are $J_r$ for  $0\le r<r_-$, $r_-\le r<r_+$, and $r\ge r_+$, respectively.
Since $V'(r)$ is continuous at the connection points, the continuity condition of the current reduces to the continuity condition of $\partial P(r,t)/\partial r$ if the radial current $J_r$ is regular at the connection points.
Using Eqs.~\eqref{current} and \eqref{P(r,t)2}, the continuity conditions can read
\begin{align}\label{connect3}
    \phi_{n}^\mathrm{in}(r_-)&=\phi_{n}^\mathrm{mid}(r_-), \quad
    \frac{d}{d r}\phi_{n}^\mathrm{in}(r)|_{r=r_-}= \frac{d}{d r}\phi_{n}^\mathrm{mid}(r)|_{r=r_-} ,\\
    \phi_{n}^\mathrm{mid}(r_+)&=\phi_{n}^\mathrm{out}\left(r_+\right), \quad
    \frac{d}{d r}\phi_{n}^\mathrm{mid}(r)|_{r=r_+}= \frac{d}{d r}\phi_{n}^\mathrm{out}(r)|_{r=r_+} ,
\label{connect4}
\end{align}
where $\phi_{n}^\mathrm{in}(r)$, $\phi_{n}^\mathrm{mid}(r)$, and $\phi_{n}^\mathrm{out}(r)$ are $\phi_{n}(r)$ for  $0\le r<r_-$, $r_-\le r<r_+$, and $r\ge r_+$, respectively.

\section{Mode analysis}\label{Sec:Mode-analysis}

\subsection{Eigenvalue equation}

Let us rewrite $\mathbb{H}$ introduced in Eq.~\eqref{mathbb_L} in polar coordinates, $r:=\sqrt{x^2+y^2}$, under the assumption of circular symmetry:
\begin{align}
    \mathbb{H}
    &= -\frac{T}{r}\frac{d}{d r}\left(r\frac{d}{d r}\right) +V_S(r) .
    \label{L_S}
\end{align}
Accordingly, we can rewrite Eq.~\eqref{Eq32} as
\begin{align}\label{eq_R}
    \phi_{n}^{\prime\prime}(r)+\frac{\phi_{n}^{\prime}(r)}{r}-\left(\beta V_S(r)-\beta \tilde{\lambda}_{n}\right)\phi_{n}(r)=0 .
\end{align}
where \(\phi_n''(r):=d^2\phi_n(r)/dr^2\), $\tilde{\lambda}_{n}:=\lambda_{n-1,0}$ with a positive integer $n=1,2, 3,\cdots$.
Let us introduce the auxiliary parameters
\begin{align}\label{nu}
    \nu_\mathrm{I} := \frac{|b_\mathrm{I}|}{2} \sqrt{\frac{\beta }{T}},\qquad
    \gamma_\mathrm{I} := \frac{\beta |k_\mathrm{I}|}{2}.
\end{align}
In this paper, we take the positive branch $\nu=|\nu|$ (choose the sign consistent with regularity at $r=0$) and $\gamma\ge0$.  

Introducing 
\begin{align}
    z_\mathrm{I}:=\gamma_\mathrm{I} r^2,\qquad 
    \Phi_{\mathrm{I},n}(z):=\phi_{\mathrm{I},n}(r)r^{-\nu_\mathrm{I}} e^{z_\mathrm{I}/2},
\end{align}
where $\Phi_{\mathrm{I},n}(z_\mathrm{I})$ satisfies Kummer's confluent-hypergeometric equation (see Appendix \ref{app:hypergeometric})
\begin{align}\label{Kummer's eq.}
    z\Phi''_{\mathrm{I},n}(z) + (1+\nu_\mathrm{I} - z)\Phi'_{\mathrm{I},n}(z) - \mu_{\mathrm{I},n}\Phi_{\mathrm{I},n}(z) = 0,
\end{align}
with the parameter
\begin{align}\label{mu}
    \mu_{\mathrm{I},n} 
    := \frac{1+\nu_\mathrm{I}}{2} 
    - \frac{\beta\left(\lambda_n + k_\mathrm{I} - \tfrac{k_\mathrm{I}|b_\mathrm{I}|}{2T}\right)}{4\gamma_\mathrm{I}}.
\end{align}

Two independent solutions of the confluent-hypergeometric equation are $M(\mu;1+\nu;z):= {}_1F_1(\mu;1+\nu;z)$ and $U(\mu;1+\nu;z)$ defined in Eq.~\eqref{U_connection} (see Appendix~\ref{app:hypergeometric} for the definitions of them and their properties).  
Hence, the general solution of Eq.~\eqref{eq_R} in the region can be expressed as
\begin{align}
    \phi_{\mathrm{I},n}(r)=r^{\nu_\mathrm{I}} e^{-\tfrac{\gamma_\mathrm{I} r^2}{2}}
    \left[ A_\mathrm{I} M\left(\mu_{\mathrm{I},n};1+\nu_\mathrm{I};\gamma_\mathrm{I} r^2\right)
    + B_\mathrm{I} U \left(\mu_{\mathrm{I},n};1+\nu_\mathrm{I};\gamma_\mathrm{I} r^2\right)\right],
\end{align} 
with constants $A_\mathrm{I}$ and $B_\mathrm{I}$ to be fixed by boundary/matching conditions.
See the explicit solution of Schr\"{o}dinger equation in Appendix \ref{app:Schroedinger}.

\section{Demonstration of the Mpemba effect}\label{Sec:Demo-Mpemba}

In this section, we examine whether the Mpemba effect can be observed.
The parameter values used throughout this section are summarized in Table~\ref{tab:parameter_sets}.
In this section, we first focus on the case for (i) $V(\alpha)<0$ in Sec. \ref{sec:V(a)<0}. 
Then, we analyze the case for (ii) $V(\alpha)>0$ in Sec. \ref{sec:V(a)>0}. 
We separately discuss the marginal case for $V(\alpha)=0$ in Appendix \ref{app:V(a)=0}.
In Sec. \ref{sec:phase_diagram}, we draw a phase diagram spanned on the parameter space controlled by $\alpha$ and $k_\mathrm{mid}$ corresponding to Fig. \ref{fig:marginal}, which covers all cases of $V(\alpha)$.

\begin{table}[htbp]
    \caption{Parameter sets used in Sec.~\ref{Sec:Demo-Mpemba}.}
    \label{tab:parameter_sets}
    \begin{ruledtabular}
    \begin{tabular}{lcccccc}
    Case &
    $T$ &
    $k_{\rm in}$ &
    $k_{\rm mid}$ &
    $k_{\rm out}$ &
    $\xi$ &
    $\alpha$
    \\ \hline
    (i) $V(\alpha)<0$
    &
    1 &
    1.0 &
    $-0.5$ &
    0.8 &
    1.0 &
    3.0
    \\
    (ii) $V(\alpha)>0$
    &
    1 &
    1.5 &
    $-1.5$ &
    1.5 &
    1.0 &
    1.5
    \\
    (iii) $V(\alpha)=0$
    &
    1 &
    1.0 &
    $-1.0$ &
    1.0 &
    1.0 &
    1.84485714
    \end{tabular}
    \end{ruledtabular}
\end{table}

First, we examine whether $a_2$ has a minimum or a maximum with respect to $\beta_\mathrm{ini}$, which is regarded as a necessary condition to observe the Mpemba effect.
We also show the behavior of $a_m$ for $m\ge 3$ against $\beta_\mathrm{ini}$.

\subsection{(i) The case for $V(\alpha)<0$}\label{sec:V(a)<0}
\subsubsection{Eigenvalues and eigenfunctions}

\begin{figure}[htbp]
    \begin{center}
    \includegraphics[height=0.3\linewidth]{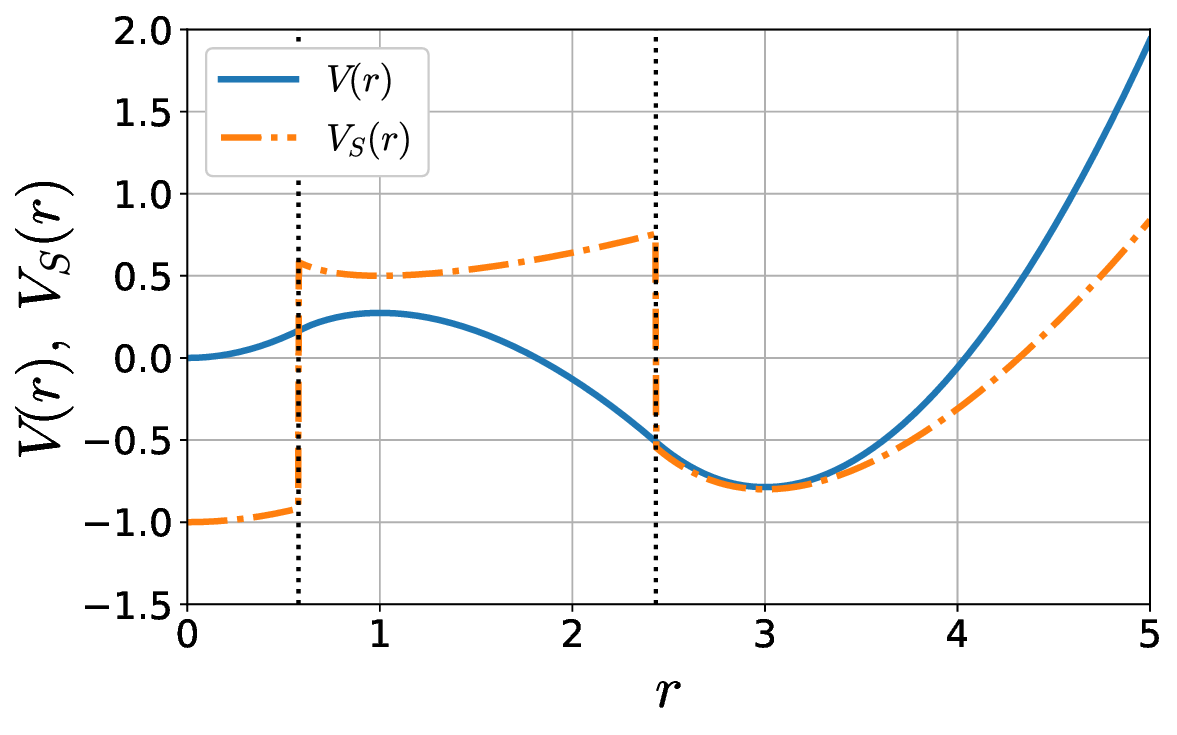}
    \end{center}
    \caption{Plots of $V(r)$ and $V_S(r)$ as functions of $r$ for the set of parameters in Eq.~\eqref{eq:parameters}.
    }
    \label{fig:potential}
\end{figure}

First, we examine the case for $V(\alpha)<0$. Thus, we adopt the parameters
\begin{equation}
    T=\beta=1,\
    k_\mathrm{in}=1.0,\ k_\mathrm{mid}=-0.5,\ k_\mathrm{out}=0.8,\  
    \xi=1.0,\ \alpha=3.0.
    \label{eq:parameters}
\end{equation}

Substituting these parameters into Eqs.~\eqref{eq:r_-_r_+} and \eqref{eq:bin_bmid_bout}, one gets
\begin{align}\label{other_parameters_i}
    \ b_\mathrm{mid}&=0.5,\ b_\mathrm{out}=-7.2,\ 
    r_- = \frac{1}{\sqrt{3}},\ 
    r_+ = \sqrt{\frac{7.7}{1.3}}\approx 2.4337372338,\\
    \mu_{\mathrm{in},n}&= \tilde{\lambda}_n,\ 
    \mu_{\mathrm{mid},n}=\tilde{\lambda}_n+\frac{5}{4},\ 
    \mu_{\mathrm{out},n}=\frac{5}{8}\tilde{\lambda}_n+\frac{18}{5}
\end{align}
Figure \ref{fig:potential} shows $V(r)$ and $V_S(r)$ for Eq.~\eqref{eq:parameters}.

\begin{figure}[htbp]
    \begin{center}
    \includegraphics[height=0.35\linewidth]{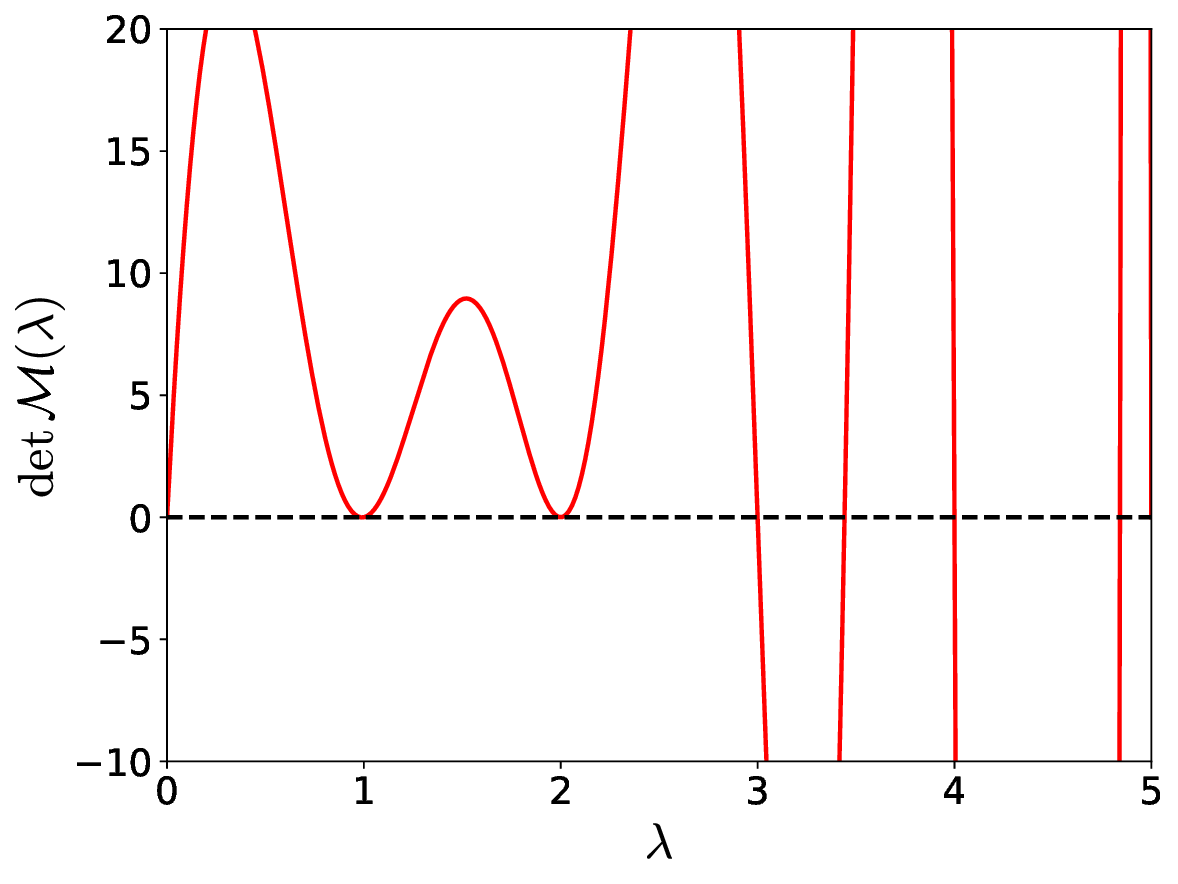}
    \end{center}
    \caption{Plot of $\det \mathcal{M}(\lambda)$ against $\lambda$ using Eq.~\eqref{eq:parameters} for $T=1$, $k_\mathrm{in}=1.0$, $k_\mathrm{mid}=-0.5$, $k_\mathrm{out}=0.8$, $\xi=1.0$, and $\alpha=3.0$.}
    \label{fig:detM}
\end{figure}

Figure \ref{fig:detM} represents the values of $\det \mathcal{M}(\lambda)$ against $\lambda$ using Eqs.~\eqref{eq:detM} and \eqref{eq:parameters} for $T=1$, $k_\mathrm{in}=1.0$, $k_\mathrm{mid}=-0.5$, $k_\mathrm{out}=0.8$, $\xi=1.0$, and $\alpha=3.0$.
It exhibits multiple solutions of $\det(\mathcal{M}(\lambda))=0$.
Interestingly, odd crossing points with $\partial_\lambda \det(\mathcal{M}(\lambda))>0$ are expressed as integers, i.e., $\lambda=0,1,2,3,4,\cdots$.
However, these zeros cannot be the eigenvalues of Eq.~\eqref{eq_R} because they satisfy another condition 
\begin{align}
    \mu_\mathrm{I,n} = -k_\mathrm{I}\quad(k_\mathrm{I}\in\mathbb{Z}_{\ge0}),
\end{align}
which leads to overcomplete conditions in Eq.~\eqref{eq:detM}.

Taking into account the normalization $\int d\bm{r} \phi_m(r)^2=1$ for arbitrary non-negative integer $m$, i.e.,
\begin{equation}\label{eq:normalization}
    \int_0^\infty dr r \phi_{m}(r)^2
    = \frac{1}{2\pi},
\end{equation}
we obtain the eigenvalues $\tilde{\lambda}_m$ and the coefficients $(A_{\mathrm{in},m},\ A_{\mathrm{mid},m},\     B_{\mathrm{mid},m},\ B_{\mathrm{out},m})$ for $2\le m \le 11$ as listed in Table \ref{table:AB}.
Substituting these parameters into Eq.~\eqref{eq:sol_R}, we can determine the eigenfunction explicitly.

\begin{table}[htbp]
    \centering
    \caption{List of values of $\tilde{\lambda}_m,\ A_{\mathrm{in},m},\ A_{\mathrm{mid},m},\     B_{\mathrm{mid},m},\ B_{\mathrm{out},m}$ for $2\le m \le 11$.}
    \begin{tabular}{c|c|c|c|c|c}
        \hline
        $m$ & $\tilde{\lambda}_m$ & $A_{\mathrm{in},m}$ & $A_{\mathrm{mid},m}$ & $B_{\mathrm{mid},m}$ & $B_{\mathrm{out},m}$ \\
        \hline
        $2$ & $9.812027\times 10^{-1}$ & $4.137916\times10^{-1}$ & $-2.929042\times10^0$ & $3.132354\times10^0$ & $-7.945768\times10^{-3}$\\
        \hline
        $3$ & $1.999213\times10^0$ & $3.456650\times10^{-1}$ & $-7.537959\times10^1$ & $-6.051495\times10^1$ & $6.254304\times10^{-3}$\\
        \hline
        $4$ & $3.441568\times10^0$ & $3.421103\times10^{-1}$ & $3.107772\times10^{-1}$ & $-4.560445\times10^{-2}$ & $-1.997224\times10^{-3}$\\
        \hline
        $5$ & $4.840318\times10^0$ & $3.698403\times10^{-1}$ & $-9.094487\times10^{-2}$ & $1.574637\times10^{-2}$ & $5.272728\times10^{-4}$\\
        \hline
        $6$ & $6.283511\times10^0$ & $3.531231\times10^{-1}$ & $5.191802\times10^{-1}$ & $8.522370\times10^{-4}$ & $-1.170742\times10^{-4}$\\
        \hline
        $7$ & $7.807206\times10^0$ & $3.352071\times10^{-1}$ & $3.261959\times10^{-2}$ & $-6.006613\times10^{-5}$ & $2.017961\times10^{-5}$\\
        \hline
        $8$ & $9.357565\times10^0$ & $3.364537\times10^{-1}$ & $4.607115\times10^{-1}$ & $-1.422739\times10^{-6}$ & $-2.886199\times10^{-6}$\\   
        \hline
        $9$ & $1.090153\times10^1$ & $3.481042\times10^{-1}$ & $-3.055038\times10^{-1}$ & $1.264037\times10^{-7}$ & $3.690271\times10^{-7}$\\
        \hline
        $10$ & $1.243967\times10^1$ & $3.601623\times10^{-1}$ & $4.515275\times10^{-1}$ & $9.513415\times10^{-10}$ & $-4.354139\times10^{-8}$\\
        \hline
        $11$ & $1.398647\times10^1$ & $3.767746\times10^
        {-1}$ & $-5.089166\times10^0$  & $-4.341282\times10^{-10}$ & $4.794075\times10^{-9}$\\
        \hline
    \end{tabular}
    \label{table:AB}
\end{table}

\subsubsection{$a_m$ for $m=2,3,4,5$}
\begin{figure}[htbp]
    \centering
    \includegraphics[width=\linewidth]{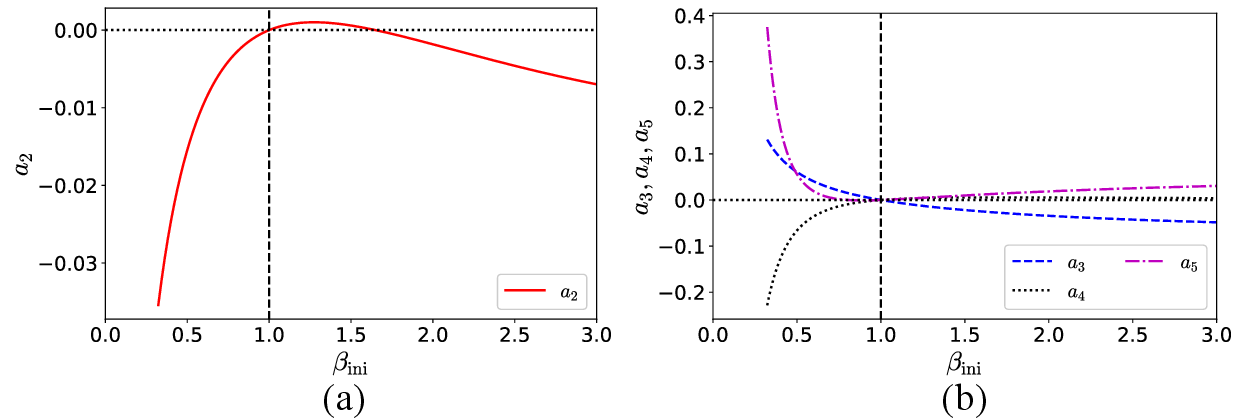}
    \caption{Plots of (a) $a_{2}(\beta_\mathrm{ini}, \beta)$ and (b) $a_{m}(\beta_\mathrm{ini}, \beta)$ ($m=3, 4,5$) against the inverse temperature $\beta_\mathrm{ini}$ for $T=1$, $k_\mathrm{in}=1.0$, $k_\mathrm{mid}=-0.5$, $k_\mathrm{out}=0.8$, $\xi=1.0$, and $\alpha=3.0$.
    Here, the vertical dashed and horizontal dotted lines in each panel correspond to $\beta_\mathrm{ini}=1$ and $a_m=0$ ($m=2, 3, 4, 5$), respectively.
    }
    \label{fig:am0}
\end{figure}

Figure \ref{fig:am0}(a) shows $\beta_\mathrm{ini}$ dependence of $a_2$ and Fig.~\ref{fig:am0}(b) shows $a_3$, $a_4$, and $a_5$. 
The results clearly demonstrate the existence of a peak of $a_2$ as a function of $\beta_\mathrm{ini}$, which is the necessary condition to observe the Mpemba effect.
Remarkably, this suggests that the overdamped Langevin system confined in a two-dimensional asymmetric potential exhibits such a peak in $a_2$ even without the presence of a hard wall.
This result contrasts with that in the one-dimensional asymmetric potential, where the Mpemba effect disappears if the system contains no hard walls~\cite{Liu26,Liu26long}.

\subsubsection{Higher-order contributions in the spectrum decomposition}

So far, we have assumed that the time evolution of the distribution function can be described by the lowest few eigenmodes in Eq.~\eqref{P(r,t)2}.
This is true if the Mpemba effect is dominated by the late-stage dynamics, but the initial relaxation might be important.

To examine the assumption that the initial relaxation is irrelevant, we examine
\begin{align}\label{P_truncated}
    P(r,t)
    &= P_\mathrm{eq}(r,\beta)
    + e^{-\beta V(r)/2} 
    \sum_{m=2}^{N_\mathrm{tr}} a_{m}\phi_{m} 
    e^{-\tilde{\lambda}_{m}t},
\end{align}
where $N_\mathrm{tr}$ is the truncation of the spectrum decomposition.
The initial distribution $P_\mathrm{ini}$ can be approximated as
\begin{align}\label{P_truncated_ini}
    P(r,0)
    &= P_\mathrm{eq}(r,\beta)
    + e^{-\beta V(r)/2} 
    \sum_{m=2}^{N_\mathrm{tr}} a_{m}\phi_{m}.
\end{align}

\begin{figure}[htbp]
    \centering
    \includegraphics[width=\linewidth]{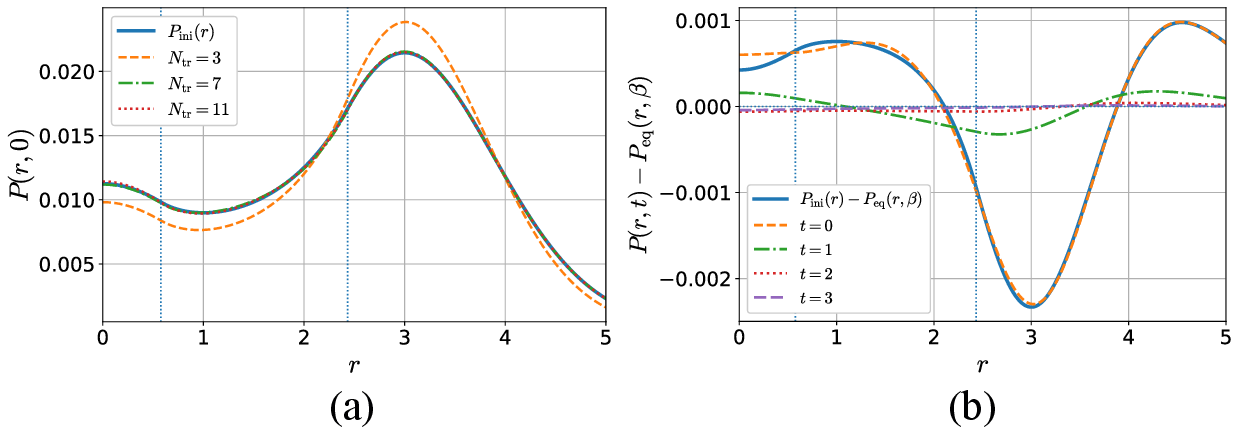}
    \caption{
    (a) Comparison between the initial distribution $P_{\rm ini}(r)$ and its truncated spectral representations for $N_{\rm tr}=3$, $7$, and $11$ with $\beta_{\rm ini}=0.82$ and $\beta=1$.
    (b) Time evolution of the difference $P(r,t)-P_{\rm eq}(r,\beta)$ for $N_{\rm tr}=11$.
    The exact initial deviation $P_{\rm ini}(r)-P_{\rm eq}(r,\beta)$ is also shown for comparison.
    We adopt Eq.~\eqref{eq:parameters} for the set of parameters.
    }
    \label{fig:N_tr}
\end{figure}

Figure \ref{fig:N_tr}(a) compares the initial distribution $P_{\rm ini}(r)$ with the truncated spectral representation in Eq.~\eqref{P_truncated_ini} for $N_{\rm tr}=3$, $7$, and $11$.
The truncated distribution approaches $P_{\rm ini}(r)$ as $N_{\rm tr}$ is increased, and the results for $N_{\rm tr}=7$ and $11$ are almost indistinguishable from the initial distribution.
This confirms the convergence of the spectral representation of the initial condition.
Figure \ref{fig:N_tr}(b) shows the time evolution of the distribution for $N_{\rm tr}=11$.
For clarity, we plot the deviation from the final equilibrium distribution, $P(r,t)-P_{\rm eq}(r,\beta)$.
The deviation progressively decays with time and approaches zero, consistent with relaxation toward $P_{\rm eq}(r,\beta)$.
The slowest nonstationary eigenvalue is $\tilde{\lambda}_2=9.812027\times10^{-1}$, corresponding to the longest relaxation time $1/\tilde{\lambda}_2\simeq1.02$.
Accordingly, the deviation becomes very small after several relaxation
times.

\subsubsection{Condition of the crossing of the KL divergence}\label{Sec:Cross_KL}

Although we have confirmed the existence of a peak of $a_2$ as a function of $\beta_\mathrm{ini}$, this may not be sufficient to observe the Mpemba effect.
To observe the Mpemba effect, we need a crossing condition for the observable.
In this subsection, we clarify the sufficient condition to observe the Mpemba effect.

Among many possible observables, we focus on the KL divergence defined as
\begin{align}\label{eq:D_KL}
    D_\mathrm{KL}(P(\beta,\beta_\mathrm{ini};t)||P_\mathrm{eq}(\beta))
    &:=\int d\bm{r} P(\bm{r},t)\ln \frac{P(\bm{r},t)}{P_\mathrm{eq}(r,\beta)} 
    =2\pi \sum_{n=2}^\infty \frac{(-1)^n}{n(n-1)}\int_0^\infty dr r \frac{(\delta P)^n}{ P_\mathrm{eq}^{n-1}},
\end{align}
which is a typical monotonic measure, where $\delta P:=P(\bm{r},t)-P_\mathrm{eq}(\bm{r})$.
If we are interested in the late stage of relaxation dynamics or the deviation from the initial state and the final state is small, $D_\mathrm{KL}(P(t)||P_\mathrm{eq})$ can be approximated as    
\begin{align}        
    D_\mathrm{KL}(P(\beta,\beta_\mathrm{ini};t)||P_\mathrm{eq}(\beta))
    &= 2\pi\int_0^\infty dr r \frac{(\delta P)^2}{2P_\mathrm{eq}}+O(\delta P^3)
    \approx
    \frac{1}{2}
    \sum_{n\ge 2}C_n
    \left(a_n\right)^2
    e^{-2\tilde{\lambda}_n t},
    \label{DKL}
\end{align}
where we have introduced
\begin{equation}\label{C_n}
C_n := \|\phi_n\|^2_{P_{\rm eq}^{-1}} > 0
\end{equation}
with $\|\phi_n\|^2_{P_{\rm eq}^{-1}}:=2\pi Z(\beta)\int_0^\infty dr r \phi_n(r)^2$.
Here, $C_n$ should be independent of $n$ because each mode $\phi_n(r)$ satisfies the normalization.
Thus, we denote $C_n$ as $C^*$.

If we are interested in the late-stage dynamics, the modes associated with the two slowest modes $\tilde{\lambda}_2$ and $\tilde{\lambda}_3$ are dominant because $e^{-\tilde{\lambda}_m t}\ll 1$ for $m\ge 4$ is negligible.
Therefore, we adopt the two-mode approximation as
\begin{align}\label{2-mode}
    D_\mathrm{KL}(P(t)||P_\mathrm{eq})\approx \frac{C^*}{2}(a_2^2 e^{-2\tilde{\lambda}_2 t}+a_3^2 e^{-2\tilde{\lambda}_3 t}).
\end{align}
Thus, the condition for a crossing of the KL divergence, characterized by Eq.~\eqref{2-mode}, is expressed as
\begin{equation}
    \mathcal{F}(\beta_\mathrm{ini}^\sharp,\beta_\mathrm{ini}^*)
    :=\frac{\Delta_3}{\Delta_2}<-1,
    \label{A4}
\end{equation}
where
\begin{equation}
    \Delta_n
    := \left(a_n^{(\mathrm{A})}\right)^2
    - \left(a_n^{(\mathrm{B})}\right)^2.
\end{equation}
for two different initial conditions A and B.
See Appendix \ref{App:crossing} for the derivation of Eq.~\eqref{A4}.
This quadratic form is valid only when $P(r,t)$ is sufficiently close to equilibrium; in particular, it does not reproduce the exact value at $t=0$ when the initial condition is another equilibrium distribution.
Therefore, Eq.~\eqref{A4} provides a sufficient condition for a crossing within the validity of the two-mode approximation, although it is not a necessary condition for the full multi-mode dynamics.

The peak of $a_{2}$ against $\beta_\mathrm{ini}$ is located around $\beta_\mathrm{ini}^*\approx 1.273$.
Let us choose another initial condition:
\begin{equation}
    \beta_\mathrm{ini}^\sharp=1.646.
    \label{eq:beta_ini_case1}
\end{equation}
In this case, we evaluate
\begin{align}
    a_{2}(\beta_\mathrm{ini}^{\sharp})
    &\approx 1.447\times10^{-5},\ 
    a_{2}(\beta_\mathrm{ini}^*)
    \approx 1.010\times10^{-3},\nonumber\\
    a_{3}(\beta_\mathrm{ini}^{\sharp})
    &\approx 5.589\times10^{-3}, \
    a_{3}(\beta_\mathrm{ini}^*)
    \approx 4.404\times10^{-3}. 
\end{align}
Then, we obtain
\begin{align}
    \Delta_2
    &:=a_{2}(\beta_\mathrm{ini}^\sharp)^2
    -a_{2}(\beta_\mathrm{ini}^{\mathrm{*}})^2
    \approx -1.021\times10^{-6}<0, \\
    \Delta_3
    &:=a_{3}(\beta_\mathrm{ini}^{\sharp})^2
    -a_{3}(\beta_\mathrm{ini}^\mathrm{*})^2
    \approx 1.184\times10^{-5}>0.
\end{align}

\begin{figure}[htbp]
    \centering
    \includegraphics[width=0.5\linewidth]{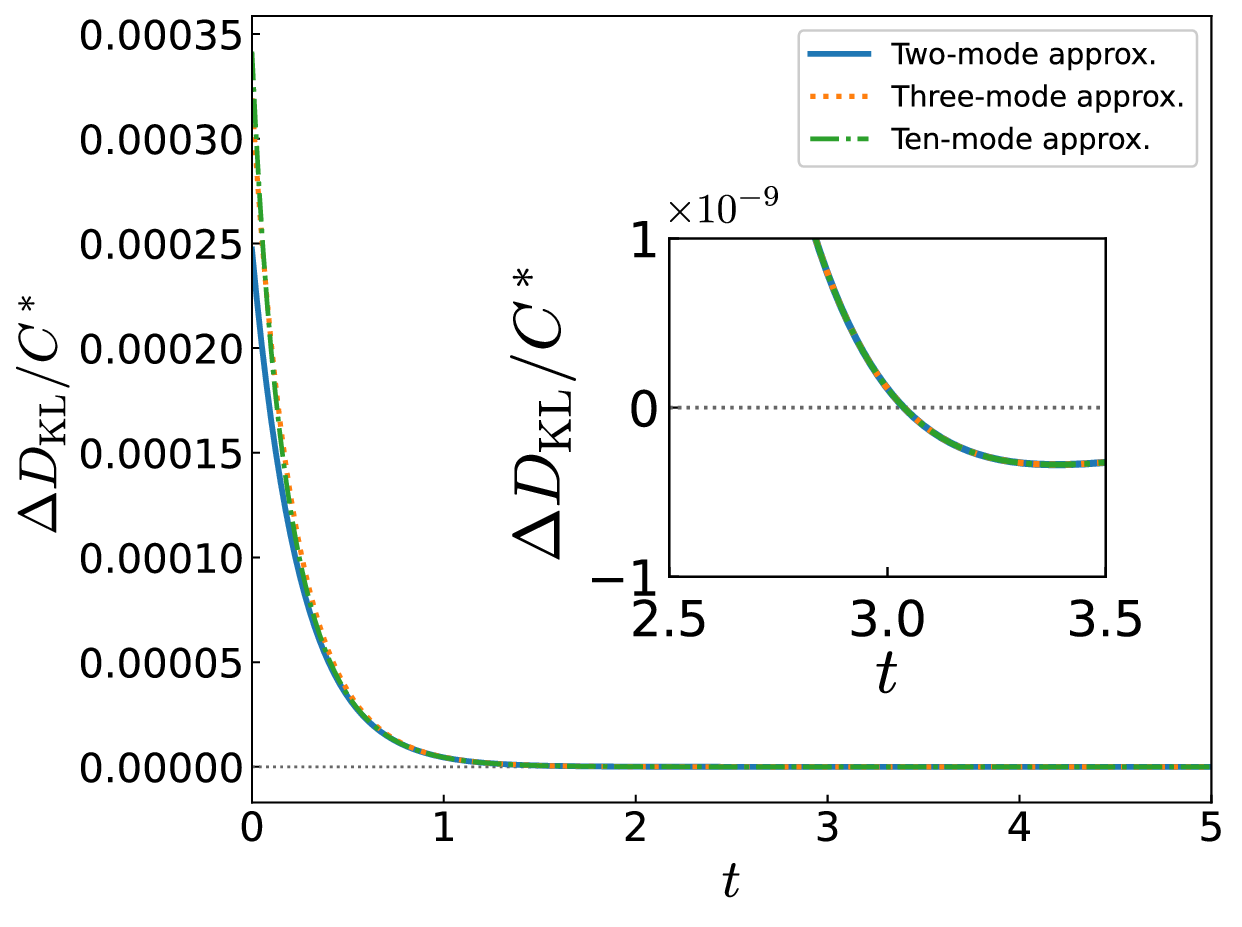}
    \caption{The demonstration of the Mpemba effect through the time evolution of the difference of the KL divergence $\Delta D:=D_\mathrm{KL}(P(\beta_\mathrm{ini}^\sharp, \beta;t)||P_\mathrm{eq}(\beta))-D_\mathrm{KL}(P(\beta_\mathrm{ini}^*, \beta;t)||P_\mathrm{eq}(\beta))$ using Eq.~\eqref{2-mode} (solid line) and Eq.~\eqref{Eq:C4} (dotted line), where we use Eq.~\eqref{eq:parameters} for the set of parameters.
    We also plot the dashed line, which includes the higher-order contributions as in Eq.~\eqref{P_truncated} with $N_\mathrm{tr}=11$.
    The inset shows the zoom around $t\simeq 3$, at which the crossing occurs.
    }
    \label{fig:KL-divergence}
\end{figure}

Furthermore, as shown in Appendix \ref{App:Higher}, the Mpemba effect persists in terms of the crossing of the KL divergences even when higher-order modes are included.
These results confirm that the truncated spectral representation provides an accurate description of both the initial distribution and its subsequent relaxation dynamics.

We plot the time evolution of the KL divergence $\Delta D(\beta_\mathrm{ini}^\sharp,\beta_\mathrm{ini}^*):=D_\mathrm{KL}(P(\beta_\mathrm{ini}^\sharp, \beta;t)||P_\mathrm{eq}(\beta))-D_\mathrm{KL}(P(\beta_\mathrm{ini}^*, \beta;t)||P_\mathrm{eq}(\beta))$ during a cooling process starting from $\beta_\mathrm{ini}^\sharp=1.646$ and $\beta_\mathrm{ini}^*=1.273$ to $\beta=1$ in Fig.~\ref{fig:KL-divergence} (see the solid line).
A crossing of the KL divergence, the Mpemba effect, is observed around $t\approx 3$ as illustrated in the inset of Fig. \ref{fig:KL-divergence}.
We now compare the KL divergence, including higher-order terms, by using the expansion of $P(r,t)$ truncated at $N_\mathrm{tr}=11$ as the dashed line.
The dotted line represents the semi-analytic result of Eq.~\eqref{Eq:C4} corresponding to $N_\mathrm{tr}=3$ (see Appendix \ref{App:Higher}). 
Thus, we confirm that the two-mode approximation is sufficient to detect the Mpemba effect, although the initial relaxation depends a little on $N_\mathrm{tr}$.
Therefore, we conclude that the crossing condition Eq.~\eqref{A4} can be used for the detection of the crossing of the KL divergence.

\begin{figure}[htbp]
    \centering
    \includegraphics[width=0.5\linewidth]{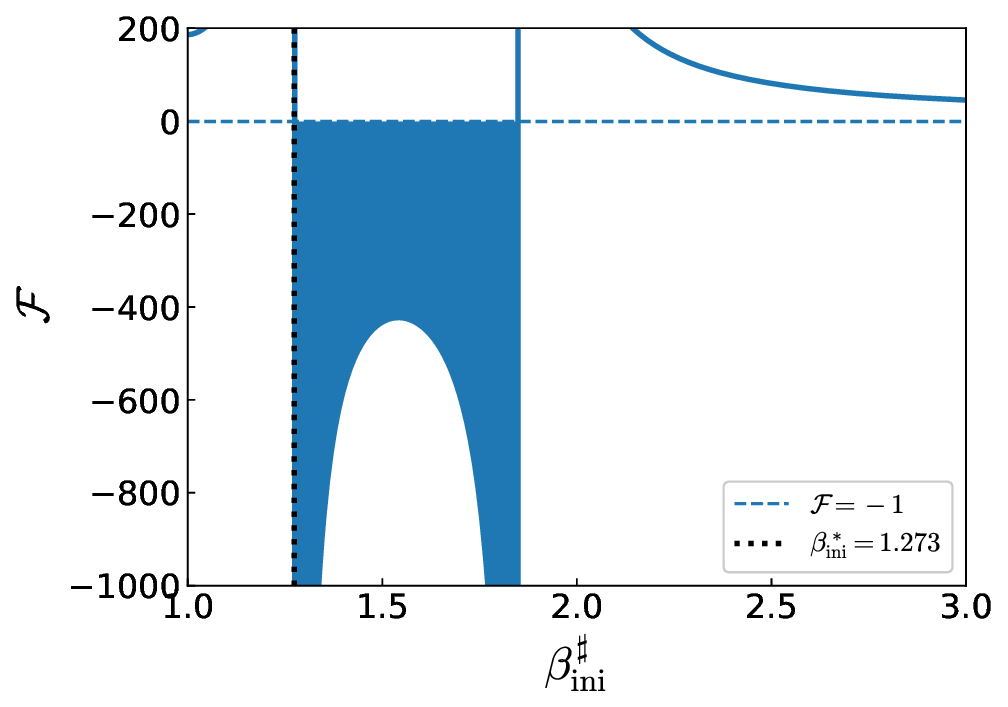}
    \caption{Plot of the KL-divergence crossing function $\mathcal{F}(\beta_\mathrm{ini}^\sharp,\beta_\mathrm{ini}^*):=\Delta_3/\Delta_2$ introduced in Eq. ~\eqref{A4} against the initial temperature $\beta_\mathrm{ini}^\sharp$, where one initial temperature is fixed at $\beta^*_\mathrm{ini}=1.273$ (the vertical dotted line), and we use Eq.~\eqref{eq:parameters} for the parameters.
    The shaded region indicates where the Mpemba effect can be observed as in Eq.~\eqref{A4}.
    }
    \label{fig:phase_diagram_case1}
\end{figure}

We draw the phase diagram to clarify the region in which the Mpemba effect (shaded region) can be observed in Fig.~\ref{fig:phase_diagram_case1} for the set of parameters in Eq.~\eqref{eq:parameters}, when we fix one initial temperature as $\beta_\mathrm{ini}=\beta_\mathrm{ini}^*$.
Since we are only interested in the Mpemba effect, the line at $\beta_\mathrm{ini}^\sharp=\beta_\mathrm{ini}^*=1.273$ is the lower bound.
Remarkably, the region where we can observe the Mpemba effect is bounded by $\beta_\mathrm{ini}^\sharp\le 1.847746\cdots$, at which $\Delta_2$ becomes zero.

We have also examined the marginal case $V(\alpha)=0$ in detail.
Although the inverse Mpemba effect is absent for the representative
parameter set, a systematic parameter scan reveals localized regions
where it can occur.
The detailed analysis is presented in Appendix~\ref{marginal}.

\subsection{Case (ii) $V(\alpha)>0$}\label{sec:V(a)>0}

\begin{figure}[htbp]
    \centering
    \includegraphics[width=\linewidth]{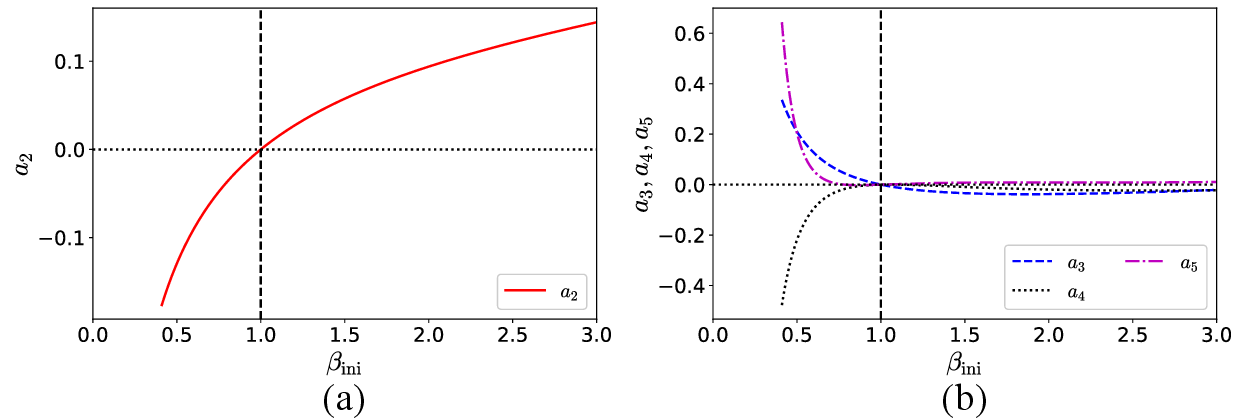}
    \caption{Plots of (a) $a_{2}(\beta_\mathrm{ini}, \beta)$ and (b) $a_{m}(\beta_\mathrm{ini}, \beta)$ ($m=3, 4,5$) against the inverse temperature $\beta_\mathrm{ini}$ when we choose the set of parameters Eq.~\eqref{eq:parameters_case3}.
    Here, the vertical dashed and horizontal dotted lines in each panel correspond to $\beta_\mathrm{ini}=1$ and $a_m=0$ ($m=2, 3, 4, 5$), respectively.
    }
    \label{fig:am0_case3}
\end{figure}

Next, let us consider case (ii) $V(\alpha)>0$.
Here, we adopt the parameters
\begin{equation}
    T=1,\
    k_\mathrm{in}=k_\mathrm{out}=1.5,\  
    \xi=1.0,\ -k_\mathrm{mid}=\alpha=1.5, 
    \label{eq:parameters_case3}
\end{equation}
as a typical example.
In this case, $a_{2}(\beta_\mathrm{ini}, \beta)$ decreases with $\beta_\mathrm{ini}$ monotonically as shown in Fig.~\ref{fig:am0_case3}(a) while the behaviors of $a_{m}(\beta_\mathrm{ini}, \beta)$ for $m=2, 3, 4$ are similar to the previous cases (see Fig. \ref{fig:am0_case3}(b)).

\subsection{Phase diagram of the Mpemba effect}\label{sec:phase_diagram}

\begin{figure}[htbp]
    \centering
    \includegraphics[width=0.5\linewidth]{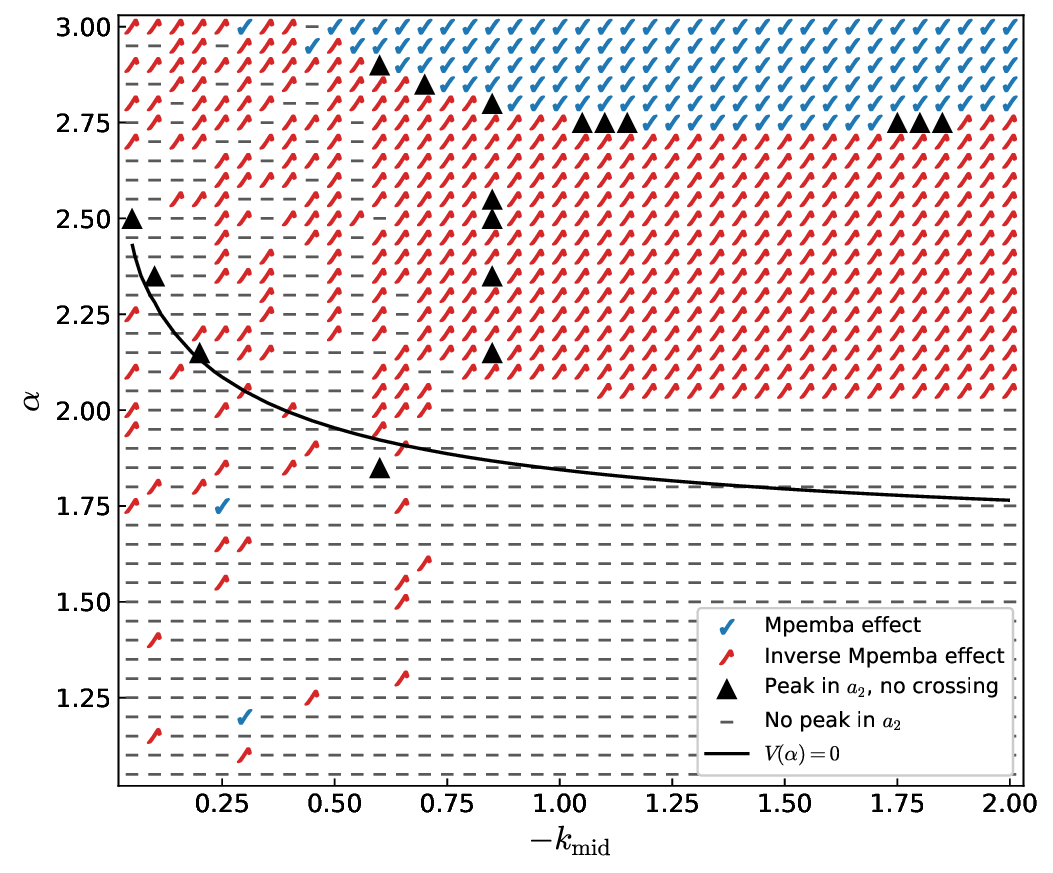}
    \caption{
    Phase diagram of the Mpemba effect in the $(-k_{\rm mid},\alpha)$ plane for $k_{\rm in}=k_{\rm out}=\xi=1$.
    The blue check marks indicate the parameter sets for which the Mpemba effect is observed, while the red inverted check marks indicate those for which the inverse Mpemba effect is observed.
    The black triangles indicate the cases in which $a_2(\beta_{\rm ini})$ exhibits a peak but the crossing condition in Eq.~\eqref{A4} is not satisfied,
    whereas the gray horizontal bars indicate the cases in which $a_2(\beta_{\rm ini})$ has no peak.
    The solid black line represents the marginal condition $V(\alpha)=0$.
     }
    \label{fig:phase_diagram}
\end{figure}

Finally, we examine the occurrence of the Mpemba effect over a wider range of the model parameters.
For each parameter set in which $a_2(\beta_{\rm ini})$ exhibits a peak, we choose $\beta_{\rm ini}^*$ at the peak position.
We then vary the other initial inverse temperature $\beta_{\rm ini}^\sharp$ over the range $0<\beta_{\rm ini}^\sharp<3$ and examine whether there exists a pair $(\beta_{\rm ini}^\sharp,\beta_{\rm ini}^*)$ that leads to a crossing of the KL divergence.
Figure~\ref{fig:phase_diagram} shows the resulting phase diagram in the $(-k_{\rm mid},\alpha)$ plane for fixed $k_{\rm in}=k_{\rm out}=\xi=1$.
The solid black line represents the marginal condition $V(\alpha)=0$ introduced in Fig.~\ref{fig:marginal}.

We classify the parameter space by relaxation behavior.
The blue check marks indicate the parameter sets for which the Mpemba effect after cooling is observed.
The region where the Mpemba effect can be observed is localized for large $\alpha$ (i.e., small $V(\alpha)$).
We also search for the inverse Mpemba effect after heating using the same procedure (see the red checks in the figure), which is still mainly placed in $V(\alpha)<0$, but some can be observed in $V(\alpha)>0$.
The black triangles indicate the cases in which $a_2(\beta_{\rm ini})$ exhibits a peak but no crossing of the KL divergence is found within the range of initial temperatures examined.
Although the observations of the black triangles are rare, we find some black triangles mainly on the boundaries $V(\alpha)=0$ or on the border between the Mpemba and the inverse Mpemba. 
The gray horizontal bars correspond to the cases in which $a_2(\beta_{\rm ini})$ has no peak, so that the necessary slowest-mode condition for the ordinary Mpemba effect is absent.
We have also checked that the black-triangle region is not an artifact of the two-mode approximation.
When higher-order contributions are included by truncating the spectral expansion in Eq.~\eqref{P_truncated} at $N_{\rm tr}=11$ and evaluating the KL divergence using Eq.~\eqref{eq:D_KL},
the black-triangle region persists. 
This confirms that the existence of a peak in $a_2(\beta_{\rm ini})$ provides only a necessary condition for the Mpemba effect, whereas its actual occurrence requires the crossing condition involving the contributions from multiple relaxation modes to be satisfied.
This overall tendency is consistent with the representative cases discussed above: the Mpemba effect or the inverse Mpemba effect is observed in case (i) with $V(\alpha)<0$ in Sec. \ref{sec:V(a)<0}, whereas it is absent for the particular parameter sets examined in cases (ii) $V(\alpha)>0$ in Sec. \ref{sec:V(a)>0} and $V(\alpha)=0$ in Appendix \ref{app:V(a)=0}, respectively.

It should be emphasized that the peak of $a_2(\beta_\mathrm{ini})$ alone does not guarantee the Mpemba effect.
In the region of black triangles, the slowest-mode amplitude $a_2$ exhibits the required nonmonotonicity, but the crossing condition for the KL divergence is not satisfied.
This demonstrates that the existence of a peak in $a_2$ is insufficient to determine the occurrence of the Mpemba effect.
Rather, the collective contribution of the relaxation modes must be taken into account through the crossing condition.

\section{Discussion}\label{Sec:Discussion}

We now discuss our results, focusing on the difference between one-dimensional and two-dimensional systems, and thermomajorization.

\subsection{Difference between one-dimensional and two-dimensional}

Remarkably, the Mpemba effect can be observed even in an unbounded system in a two-dimensional situation, while we need to introduce a wall to confine the particle in a one-dimensional asymmetric potential~\cite{Liu26,Liu26long}.
The difference may be due to the forbidden region for $r<0$ in two-dimensional cases, i.e., there is an effective wall at $r=0$ in the two-dimensional case.
More explicitly, the equilibrium condition for the two-dimensional system is expressed as 
\begin{align}
    P_\mathrm{eq}(r)\propto r e^{-\beta V(r)}.
\end{align}
Thus, the effective potential mapping onto one-dimensional systems can be written as
\begin{align}
    V_\mathrm{eff}(r)=V(r)-T\ln r .
\end{align}
The logarithmic term in $V_\mathrm{eff}(r)$  comes from the Jacobian.
This effective potential prohibits the location of a test particle for $r<0$, which can be regarded as an effective wall at $r=0$.
Since the effective potential in the $d$-dimensional case is given by $V_\mathrm{eff}(r)=V(r)-T(d-1)\ln r$, we expect that arbitrary systems for $d>1$ are independent of system size.
We can prove this expectation by showing that the peak of $a_2(\beta_\mathrm{ini})$ is independent of the system size for $d>1$ dimensions in Appendix \ref{app:system_size}.

The necessary condition to observe the Mpemba effect is whether 
\begin{align}\label{F_2}
    F_2(\beta_\mathrm{ini}):=\frac{\partial a_2}{\partial \beta_\mathrm{ini}}= 
    -2\pi \int_0^\infty dr r e^{\beta V(r)/2}\phi_2(r)[V(r)-\langle V\rangle_{\beta_\mathrm{ini}}]P_\mathrm{ini}(r) 
\end{align}
has, at least, one zero for $\beta_\mathrm{ini}=\beta_\mathrm{ini}^*$.
We demonstrate that $F_2(\beta_\mathrm{ini})$ has a different sign in the low-temperature and high-temperature limits for the case (i) $V(\alpha)<V(0)=0$, as shown in Appendix~\ref{App:D1}.
Thus, we conclude the existence of $F_2(\beta_\mathrm{ini}^*)=0$ at $\beta_\mathrm{ini}=\beta_\mathrm{ini}^*$.
Mathematically, the difference in the Jacobian between two-dimensional and one-dimensional leads to different conclusions, while $r=0$ can be regarded as an effective potential in the one-dimensional problem.
Interestingly, we can prove the absence of positive $\beta_\mathrm{ini}^*$ to satisfy $F_2(\beta_\mathrm{ini}^*)$ for the case (ii) $V(\alpha)>0$ (see Appendix \ref{App:D2}). 
When we regard $r=0$ as an effective potential in a one-dimensional system, this result is analogous to that in the one-dimensional case with a wall on one side~\cite{Liu26,Liu26long}.

\subsection{Relation to Thermomajorization and Measure Independence}

Recently, Vu and Hayakawa~\cite{VVu2025} established that thermomajorization provides a necessary and sufficient initial-state condition to observe the Mpemba effect across arbitrary monotone measures. While this geometric framework offers an elegant, measure-independent criterion for the thermodynamic admissibility of anomalous relaxation, constructing full thermomajorization curves can be practically challenging and does not directly yield real-time kinetic features. 

Crucially, our explicit multi-mode framework demonstrates that once the expansion coefficients $a_2$, $a_3$, and the corresponding eigenvalues are known analytically, the abstract machinery of thermomajorization is no longer strictly required to predict a crossing. Indeed, the derivation presented in Sec.~\ref{Sec:Cross_KL} can be seamlessly generalized to any arbitrary observable $\hat{A}$ via Eq.~\eqref{<A>}. Under this representation, the existence of a local extremum in the slowest-mode amplitude $a_2(\beta_{\rm ini})$ remains a universal necessary condition to observe the Mpemba effect for any observable. 

Furthermore, by explicitly evaluating $a_2$ and $a_3$, one can directly write down the exact crossing condition—analogous to Eq.~\eqref{A4}—for any chosen physical quantity or information-theoretic metric. Consequently, our explicit kinetic approach guarantees that if the multi-mode crossing condition is satisfied for a fundamental distance measure like the KL divergence, the presence of anomalous relaxation is robustly determined by the intrinsic mode dynamics, removing any ambiguity regarding whether specific monotone measures might fail to cross during the relaxation process.

\subsection{Experimental realization}

Finally, we discuss the physical and experimental realizability of the spherically symmetric, multi-well potential analyzed in this work. While our primary motivation stems from its mathematical tractability and analytical solvability, this class of potentials closely mirrors several quasi-realistic physical platforms. 
Most notably, in soft matter physics, modern optical tweezers equipped with spatial light modulators or high-speed scanning mirrors allow experimentalists to routinely sculpt highly precise, multi-stable, and concentric conservative potential landscapes for colloidal Brownian particles. Such microengineered setups have already become a premier testbed for verifying anomalous thermal relaxation and the Mpemba effect in the laboratory. 
Additionally, from a broader thermodynamic perspective, our $d$-dimensional radial potential can be interpreted as an effective free-energy landscape for homogeneous nucleation or bubble cavitation, where the reaction coordinate is intrinsically the physical radius ($r$) of the growing phase, naturally introducing the hyperspherical volume element $r^{d-1}$. Our model thus provides a directly applicable framework for predicting and controlling anomalous relaxation rates in these diverse experimental systems.

\section{Conclusion}\label{Sec:Conclusion}

In this paper, we have introduced an exactly solvable two-dimensional bistable potential and obtained exact eigenmodes of the corresponding Fokker-Planck equation. 
The spectral decomposition enables a systematic analysis of the relaxation dynamics after a quench between equilibrium states and provides a theoretical framework for understanding the Mpemba effect in terms of the relaxation modes.

Within the two-mode approximation, we derived the condition for the crossing of the KL divergence and showed that the slowest-mode amplitude alone does not determine the occurrence of the Mpemba effect. 
Instead, the crossing is governed by the interplay between the slowest and the second-slowest relaxation modes. By extending the analysis to higher-order modes, we further confirmed that the criterion obtained from the two-mode approximation remains valid, demonstrating that a peak in the slowest-mode amplitude is only a necessary condition, whereas the actual occurrence of the Mpemba effect requires the crossing condition involving the relaxation modes to be satisfied.

Applying the theory to the present model, we find a strong correlation
between the occurrence of the Mpemba effect and the relative depths of
the two minima, although the sign of $V(\alpha)$ does not provide a
strict criterion.
The Mpemba effect is predominantly observed when the outer minimum is
deeper than the central minimum, i.e., $V(\alpha)<0$.
However, the marginal condition $V(\alpha)=0$ does not by itself
preclude the Mpemba effect, although it is absent for the representative
parameter set considered in Sec.~IV D, the inverse Mpemba effect appears
in localized regions when the curvatures of the potential are varied.
Similarly, a small number of parameter sets exhibiting the Mpemba effect
are found on the $V(\alpha)>0$ side.
These results demonstrate that the relative well depths alone are
insufficient to determine the occurrence of the Mpemba effect and that
the detailed relaxation spectrum must be taken into account.

Unlike one-dimensional bistable models, the present two-dimensional system exhibits the Mpemba effect without introducing artificial confining walls, since the origin acts as an effective reflecting boundary through the radial geometry. 
The mathematical proof that the Mpemba effect is possible only for $V(\alpha)<0$ is presented in Appendix~\ref{app:F_2}. 
Physically, when $V(\alpha)<0$, the increase in the available phase-space volume at large radii competes with the energetic preference for the low-temperature state, giving rise to the nonmonotonic relaxation responsible for the Mpemba effect. 
When $V(\alpha)>0$, the energetic and geometrical effects cooperate, leading to monotonic relaxation and preventing the appearance of the Mpemba effect.

The present framework provides an analytically tractable platform for investigating anomalous relaxation phenomena in multidimensional systems. 
It would be interesting to extend the analysis to anisotropic potentials, non-equilibrium initial states, and many-body systems, where richer relaxation dynamics and additional Mpemba-like phenomena may emerge.

\section*{Acknowledgment}
We thank Y. Liu for stimulating discussion.
HH also thanks N. Ohga, F. van Wijland, R. Chetrite, and T. Van Vu for fruitful discussions.
This work is partially supported by JSPS Kakenhi (Grant Nos.~JP24K06974, JP24K07193, and JP25K01063).

\newpage

\appendix

\section{Details of the potential form}\label{app:potential}
In this Appendix, we present the explicit expressions of $r_-$, $r_+$, $C_\mathrm{mid}$, and $C_\mathrm{out}$.
We also write the explicit expression of $V(\alpha)$, which is used in the main text.

The derivatives in each region are given by
\begin{align}\label{V'(r)}
    V'_{\rm in}(r)&=k_{\rm in} r, \quad
    V'_{\rm mid}(r)=k_{\rm mid}\left(r-\frac{\xi^2}{r}\right), \quad
    V'_{\rm out}(r)=k_{\rm out}\left(r-\frac{\alpha^2}{r}\right).
\end{align}
The second derivative of $V(r)$ are given by
\begin{align}
    V''_{\rm in}(r=0)=k_{\rm in}, \quad 
    V''_{\rm mid}(r=\xi)=2k_{\rm mid}, \quad
    V''_{\rm out}(r=\alpha)=2k_{\rm out},
\end{align}
where $V''(r)=d^2V(r)/dr^2$.
The conditions for bistability are, therefore, given by
\begin{align}\label{cond_a}
    k_{\rm in}>0, \qquad k_{\rm mid}<0, \qquad k_{\rm out}>0.
\end{align}

Continuity of $V'(r)$ at $r=r_\pm$ gives explicit formulas for the matching radii:
\begin{align}
    r_- &= \xi \sqrt{\frac{-k_{\rm mid}}{k_{\rm in}-k_{\rm mid}}}, \qquad
    r_+ = \sqrt{\frac{k_{\rm out}\alpha^2-k_{\rm mid}\xi^2}{\,k_{\rm out}-k_{\rm mid}\,}} .
    \label{eq:r_-_r_+}
\end{align}
For the sign choice above, these roots are real and positive, satisfying $0<r_-<\xi<r_+<\alpha$ for appropriate parameter sets. 

Continuity of $V(r)$ at $r_\pm$ fixes $C_{\rm mid}$ and $C_{\rm out}$:
\begin{align}
    C_{\rm mid} &= \frac{k_{\rm in}-k_{\rm mid}}{2} r_-^2 - b_{\rm mid}\ln r_- + C_{\rm in}, 
    \label{C_mid}\\
    C_{\rm out} &= \frac{k_{\rm mid}-k_{\rm out}}{2} r_+^2 + (b_{\rm mid}-b_{\rm out})\ln r_+ + C_{\rm mid}.
    \label{C_out}
\end{align}
One may set $C_{\rm in}=0$ without loss of generality.

The potential $V(r)$ in Eq.~\eqref{V'(r)} has two minima at $r=0$ and $r=\alpha$.
Then, the values of $V(r)$ at these minima are, respectively, given by $V(0)=0$, and
\begin{align}
    V(\alpha)&=
   \frac{k_{\rm in}}{2} r_-^2
    +\frac{k_{\rm mid}}{2}\left(r_+^2 - r_-^2\right)
    +\frac{k_{\rm out}}{2}\left(\alpha^2 - r_+^2\right)
    + b_{\rm mid}\ln\!\left(\frac{r_+}{r_-}\right)
    + b_{\rm out}\ln\!\left(\frac{\alpha}{r_+}\right) \notag\\
    &= \frac{1}{2}k_{\rm out}\alpha^2\ln \frac{k_{\rm out}\alpha^2-k_{\rm mid}\xi^2}{\alpha^2(k_{\rm out}-k_{\rm mid})}
    + \frac{1}{2}k_{\rm mid}\xi^2\ln \left(\frac{k_{\rm out}-k_{\rm mid}}{k_{\rm in}-k_{\rm mid}}\frac{-k_{\rm mid}\xi^2}{k_{\rm out}\alpha^2-k_{\rm mid}\xi^2}\right),
\end{align}
where we have used Eqs.~\eqref{pot2}, \eqref{eq:r_-_r_+}, \eqref{C_mid}, and \eqref{C_out}.

\section{Properties of Tricomi's $U$-function}\label{app:hypergeometric}
The Tricomi (or Tricomi-Confluent) function $U(a,b,z)$ is the second, linearly independent solution of Kummer's (confluent-hypergeometric) differential equation
\begin{equation}\label{Kummer}
    z y''(z) + (b-z) y'(z) - a y(z) = 0.
\end{equation}
The other standard solution is Kummer's $M$-function (also denoted ${}_1F_1$)~\cite{Abramowitz}:
\begin{equation}\label{def:M}
    M(a,b,z):= {}_1F_1(a;b;z)
    =\sum_{k=0}^{\infty}\frac{(a)_k}{(b)_k}\frac{z^k}{k!},
\end{equation}
where $(q)_k:=\Gamma(q+k)/\Gamma(q)$  with the Gamma function $\Gamma(x):=\int_0^\infty dt t^{x-1}e^{-t}$ is the Pochhammer symbol.

A convenient and frequently used representation of $U(a,b,z)$ is the integral formula (valid for $\Re a>0,\ \Re z>0$):
\begin{equation}\label{U_integral}
    U(a,b,z) := \frac{1}{\Gamma(a)}\int_{0}^{\infty} e^{-z t} t^{a-1} (1+t)^{b-a-1}dt .
\end{equation}
By analytic continuation, this defines $U(a,b,z)$ for general complex parameters and $z$ (with the usual branch cuts).

The connection formula expressing $U$ in terms of $M$ is
\begin{equation}\label{U_connection}
    U(a,b,z)
    = \frac{\Gamma(1-b)}{\Gamma(a-b+1)} M(a,b,z)
    + \frac{\Gamma(b-1)}{\Gamma(a)} z^{1-b} M(a-b+1,2-b,z),
\end{equation}
which is valid by analytic continuation for non-integer $b$ and extends to integer $b$ by limit.
Useful identities and asymptotic form are presented as
\begin{align}
    &\frac{d}{dz}M(a,b,z)=\frac{a}{b}M(a+1,b+1,z),\\
    &\frac{d}{dz}U(a,b,z) = -a\,U(a+1,b+1,z),\label{U_derivative}\\
    &U(a,b,z) \sim z^{-a}\Big(1+O(1/z)\Big)\qquad(\lvert z\rvert\to\infty,\ \Re z>0).
\end{align}

In our application (with the notation of the main text), one sets
\begin{align}
    a=\mu,\qquad b=1+\nu,\qquad z=\gamma r^2,
\end{align}
so that $U(\mu;1+\nu;\gamma r^2)$ denotes Tricomi's function with those parameters.  
The identity \eqref{U_derivative} is particularly useful when computing derivatives of the radial solution
\begin{align}
    \phi(r)=r^{\nu}e^{-\tfrac{\gamma r^2}{2}}U(\mu,1+\nu,\gamma r^2)
\end{align}
appearing in the matching conditions.

\section{Solution of Schr\"{o}dinger equation}\label{app:Schroedinger}

\subsection{Outline of the solution of Schr\"{o}dinger equation}

First, we consider this situation.
Because a solution should not diverge at $r\to0$ and $r\to\infty$, it follows that
\begin{equation}
    \phi_n(r)=
    \begin{cases}
    A_\mathrm{in}\Phi_\mathrm{in,M}^{(1)}(r) 
    & (0\le r < r_-)\\
    A_\mathrm{mid}\Phi_\mathrm{mid,M}^{(1)}(r) 
    + B_\mathrm{mid}\Phi_\mathrm{mid,U}^{(1)}(r)
    & (r_- \le r < r_+)\\
    B_\mathrm{out}\Phi_\mathrm{out,U}^{(1)}(r)
    & (r \ge r_+)
    \end{cases},
    \label{eq:sol_R}
\end{equation}
where $\Phi_\mathrm{I,M}^{(1)}(r)$ and $\Phi_\mathrm{I,U}^{(1)}(r)$ satisfy Eq.~\eqref{Kummer's eq.}, which are, respectively, given by
\begin{equation}
    \Phi_{\mathrm{I},\mathrm{M}}^{(1)}(r) 
    := r^{\nu_\mathrm{I}}e^{-\tfrac{\gamma_\mathrm{I} r^2}{2}}
    M(\mu_\mathrm{I}; 1+\nu_\mathrm{I}; \gamma_\mathrm{I} r^2),\quad
    \Phi_{\mathrm{I},\mathrm{U}}^{(1)}(r)
    := r^{\nu_\mathrm{I}}e^{-\tfrac{\gamma_\mathrm{I} r^2}{2}}
    U(\mu_\mathrm{I}; 1+\nu_\mathrm{I}; \gamma_\mathrm{I} r^2)
\end{equation}
with Eq. \eqref{U_connection}. 

Using this solution, the linear matching conditions at $r=r_-$ and $r=r_+$ given by Eqs.~\eqref{connect3} and \eqref{connect4} become
\begin{align}
    \begin{pmatrix}
        \Phi^{(1)}_{\rm in, M}(r_-) & -\Phi^{(1)}_{\rm mid,M}(r_-) 
        & -\Phi^{(1)}_{\rm mid,U}(r_-) & 0 \\
        \Phi^{(2)}_{\rm in,M}(r_-) & -\Phi^{(2)}_{\rm mid,M}(r_-) 
        & -\Phi^{(2)}_{\rm mid,U}(r_-) & 0 \\
        0 & \Phi^{(1)}_{\rm mid,M}(r_+) 
        & \Phi^{(1)}_{\rm mid,U}(r_+) & -\Phi^{(1)}_{\rm out,U}(r_+) \\
        0 & \Phi^{(2)}_{\rm mid,M}(r_+) 
        & \Phi^{(2)}_{\rm mid,U}(r_+) & -\Phi^{(2)}_{\rm out,U}(r_+)
    \end{pmatrix}
    \begin{pmatrix}
        A_\mathrm{in} \\ A_\mathrm{mid}\\
        B_\mathrm{mid} \\ B_\mathrm{out}
    \end{pmatrix}
    = 
    \begin{pmatrix}
        0 \\ 0 \\ 0 \\ 0
    \end{pmatrix} .
    \label{eq:simul_eq}
\end{align}
where $\Phi_{\mathrm{I},\mathrm{M}}^{(2)}(r)$ and $\Phi_{\mathrm{I},\mathrm{U}}^{(2)}(r)$ are given by
\begin{align}
    \Phi_{\mathrm{I},\mathrm{M}}^{(2)}(r)
    := \frac{\partial}{\partial r}\Phi_{\mathrm{I},\mathrm{M}}^{(1)}(r),\quad
    \Phi_{\mathrm{I},\mathrm{U}}^{(2)}(r)
    := \frac{\partial}{\partial r}\Phi_{\mathrm{I},\mathrm{U}}^{(1)}(r).
\end{align}

Nontrivial solutions exist only if
\begin{equation}\label{eq:detM}
    \det\mathcal{M}(\lambda)=0 ,
\end{equation}
where the matrix $\mathcal{M}$ is defined as
\begin{equation}    
    \mathcal{M}:=
    \begin{pmatrix}
        \Phi^{(1)}_{\rm in, M}(r_-) & -\Phi^{(1)}_{\rm mid,M}(r_-) 
        & -\Phi^{(1)}_{\rm mid,U}(r_-) & 0 \\
        \Phi^{(2)}_{\rm in,M}(r_-) & -\Phi^{(2)}_{\rm mid,M}(r_-) 
        & -\Phi^{(2)}_{\rm mid,U}(r_-) & 0 \\
        0 & \Phi^{(1)}_{\rm mid,M}(r_+) 
        & \Phi^{(1)}_{\rm mid,U}(r_+) & -\Phi^{(1)}_{\rm out,U}(r_+) \\
        0 & \Phi^{(2)}_{\rm mid,M}(r_+) 
        & \Phi^{(2)}_{\rm mid,U}(r_+) & -\Phi^{(2)}_{\rm out,U}(r_+)
    \end{pmatrix}.
\end{equation}
Interestingly, the solutions of Eq.~\eqref{eq:detM} are classified into two types: trivial solutions with integer eigenvalues and nontrivial solutions. 
As will be seen, only the nontrivial solutions are relevant. 
For the explicit procedure to determine the eigenmodes, see Appendix \ref{App:det_eigen}.

\subsection{Determination of the eigenmodes}\label{App:det_eigen}

In this appendix, we describe the procedure for determining the eigenmodes.
First, we can rewrite Eq.~\eqref{eq:detM} explicitly as
\begin{align}
    \det\mathcal{M}(\lambda)
    &= \left[\Phi^{(1)}_{\rm in, M}(r_-)\Phi^{(2)}_{\rm mid,M}(r_-)
    - \Phi^{(2)}_{\rm in,M}(r_-)\Phi^{(1)}_{\rm mid,U}(r_-)\right]
    \left[\Phi^{(1)}_{\rm mid,U}(r_+)\Phi^{(2)}_{\rm out,U}(r_+)
    - \Phi^{(2)}_{\rm mid,U}(r_+)\Phi^{(1)}_{\rm out,U}(r_+)\right]\nonumber\\
    &\hspace{1em}
    + \left[\Phi^{(1)}_{\rm in, M}(r_-)\Phi^{(2)}_{\rm mid,U}(r_-)
    - \Phi^{(2)}_{\rm in,M}(r_-)\Phi^{(1)}_{\rm mid,U}(r_-)\right]
    \left[\Phi^{(2)}_{\rm mid,M}(r_+)\Phi^{(1)}_{\rm out,U}(r_+)
    - \Phi^{(1)}_{\rm mid,M}(r_+) \Phi^{(2)}_{\rm out,U}(r_+)\right].
\end{align}
This expression helps one to numerically determine $\lambda$ satisfying Eq.~\eqref{eq:detM}.

As explained in the main text, let us focus on the case where the eigenvalues are nontrivial.
Let us determine the coefficients $A_\mathrm{in}$--$B_\mathrm{out}$.
First, we consider the following set of equations:
\begin{equation}
    \begin{pmatrix}
        \Phi^{(1)}_{\rm mid,M}(r_-) 
        & \Phi^{(1)}_{\rm mid,U}(r_-)
        & 0\\
        \Phi^{(2)}_{\rm mid,M}(r_-) 
        & \Phi^{(2)}_{\rm mid,U}(r_-)
        & 0\\
        \Phi^{(1)}_{\rm mid,M}(r_+) 
        & \Phi^{(1)}_{\rm mid,U}(r_+) & -\Phi^{(1)}_{\rm out,U}(r_+) 
    \end{pmatrix}
    \begin{pmatrix}
        A_\mathrm{mid} \\ B_\mathrm{mid} \\ B_\mathrm{out}
    \end{pmatrix}
    =
    \begin{pmatrix}
        \Phi^{(1)}_{\rm in,M}(r_-)\\
        \Phi^{(2)}_{\rm in,M}(r_-)\\
        0
    \end{pmatrix}
    A_\mathrm{in}.
\end{equation}
We can easily solve this set of equations as
\begin{align}
    \frac{A_\mathrm{mid}}{A_\mathrm{in}}
    &= \frac{\Phi^{(1)}_{\rm in,M}(r_-)\Phi^{(2)}_{\rm mid,U}(r_-)-\Phi^{(2)}_{\rm in,M}(r_-)\Phi^{(1)}_{\rm mid,U}(r_-)}{\Phi^{(1)}_{\rm mid,M}(r_-)\Phi^{(2)}_{\rm mid,U}(r_-)-\Phi^{(1)}_{\rm mid,U}(r_-)\Phi^{(2)}_{\rm mid,M}(r_-)},
    \label{eq:A_mid_A_in}\\
    \frac{B_\mathrm{mid}}{A_\mathrm{in}}
    &= \frac{-\Phi^{(1)}_{\rm in,M}(r_-)\Phi^{(2)}_{\rm mid,M}(r_-)+\Phi^{(2)}_{\rm in,M}(r_-)\Phi^{(1)}_{\rm mid,M}(r_-)}{\Phi^{(1)}_{\rm mid,M}(r_-)\Phi^{(2)}_{\rm mid,U}(r_-)-\Phi^{(1)}_{\rm mid,U}(r_-)\Phi^{(2)}_{\rm mid,M}(r_-)},\\
    \frac{B_\mathrm{out}}{A_\mathrm{in}}
    &= \frac{\Phi^{(1)}_{\rm in,M}(r_-)}{\Phi^{(1)}_{\rm out,U}(r_+)}
    \frac{\Phi^{(2)}_{\rm mid,U}(r_-)\Phi^{(1)}_{\rm mid,M}(r_+)-\Phi^{(2)}_{\rm mid,M}(r_-)\Phi^{(1)}_{\rm mid,U}(r_+)}{\Phi^{(1)}_{\rm mid,M}(r_-)\Phi^{(2)}_{\rm mid,U}(r_-)-\Phi^{(1)}_{\rm mid,U}(r_-)\Phi^{(2)}_{\rm mid,M}(r_-)}\nonumber\\
    &\hspace{1em}
    +\frac{\Phi^{(2)}_{\rm in,M}(r_-)}{\Phi^{(1)}_{\rm out,U}(r_+)}
    \frac{\Phi^{(1)}_{\rm mid,M}(r_-)\Phi^{(1)}_{\rm mid,U}(r_+)-\Phi^{(1)}_{\rm mid,U}(r_-)\Phi^{(1)}_{\rm mid,M}(r_+)}{\Phi^{(1)}_{\rm mid,M}(r_-)\Phi^{(2)}_{\rm mid,U}(r_-)-\Phi^{(1)}_{\rm mid,U}(r_-)\Phi^{(2)}_{\rm mid,M}(r_-)}.
    \label{eq:B_out_A_in}
\end{align}
Combining Eqs.~\eqref{eq:A_mid_A_in}--\eqref{eq:B_out_A_in} with the normalization condition given in Eq.~\eqref{eq:normalization}, one can determine the coefficients $A_\mathrm{in}$, $A_\mathrm{mid}$, $B_\mathrm{mid}$, and $B_\mathrm{out}$.

\section{Marginal case $V(\alpha)=0$}\label{marginal}\label{app:V(a)=0}

In this Appendix, we examine the marginal case for $V(\alpha)=0$, where the two potential minima at $r=0$ and $r=\alpha$ have the same depth.
We first consider a representative parameter set and examine the relaxation-mode amplitudes and the crossing condition.
We then systematically vary the curvatures of the potential to investigate whether the inverse Mpemba effect can occur under the marginal condition $V(\alpha)=0$.

\begin{figure}[htbp]
    \centering
    \includegraphics[width=\linewidth]{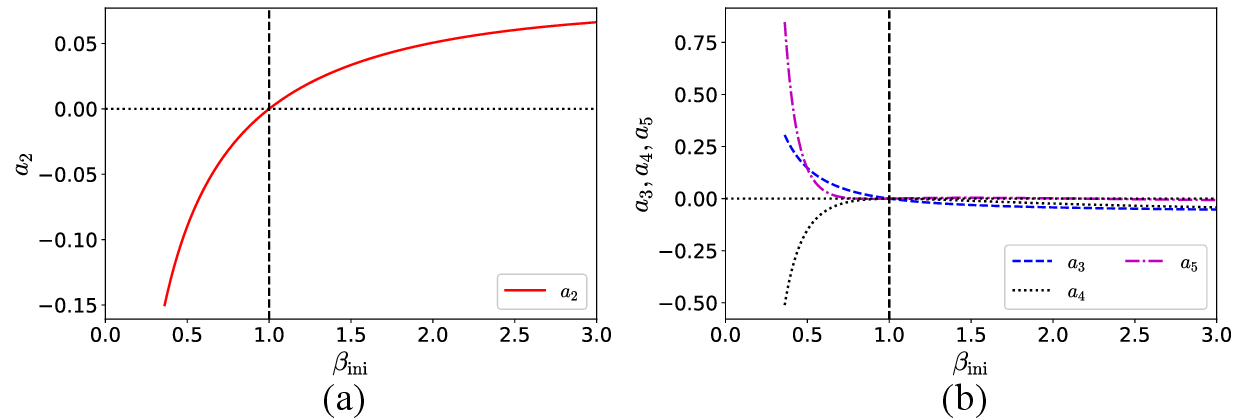}
    \caption{Plots of (a) $a_{2}(\beta_\mathrm{ini}, \beta)$ and (b) $a_{m}(\beta_\mathrm{ini}, \beta)$ ($m=3, 4,5$) against the inverse temperature $\beta_\mathrm{ini}$ for $T=1$, $k_\mathrm{in}=-k_\mathrm{mid}=k_\mathrm{out}=1$, $\xi=1.0$, and $\alpha=1.84485714$.
    Here, the vertical dashed and horizontal dotted lines in each panel correspond to $\beta_\mathrm{ini}=1$ and $a_m=0$ ($m=2, 3, 4, 5$), respectively.
    }
    \label{fig:am_case2}
\end{figure}
\begin{figure}[htbp]
    \centering
    \includegraphics[width=0.5\linewidth]{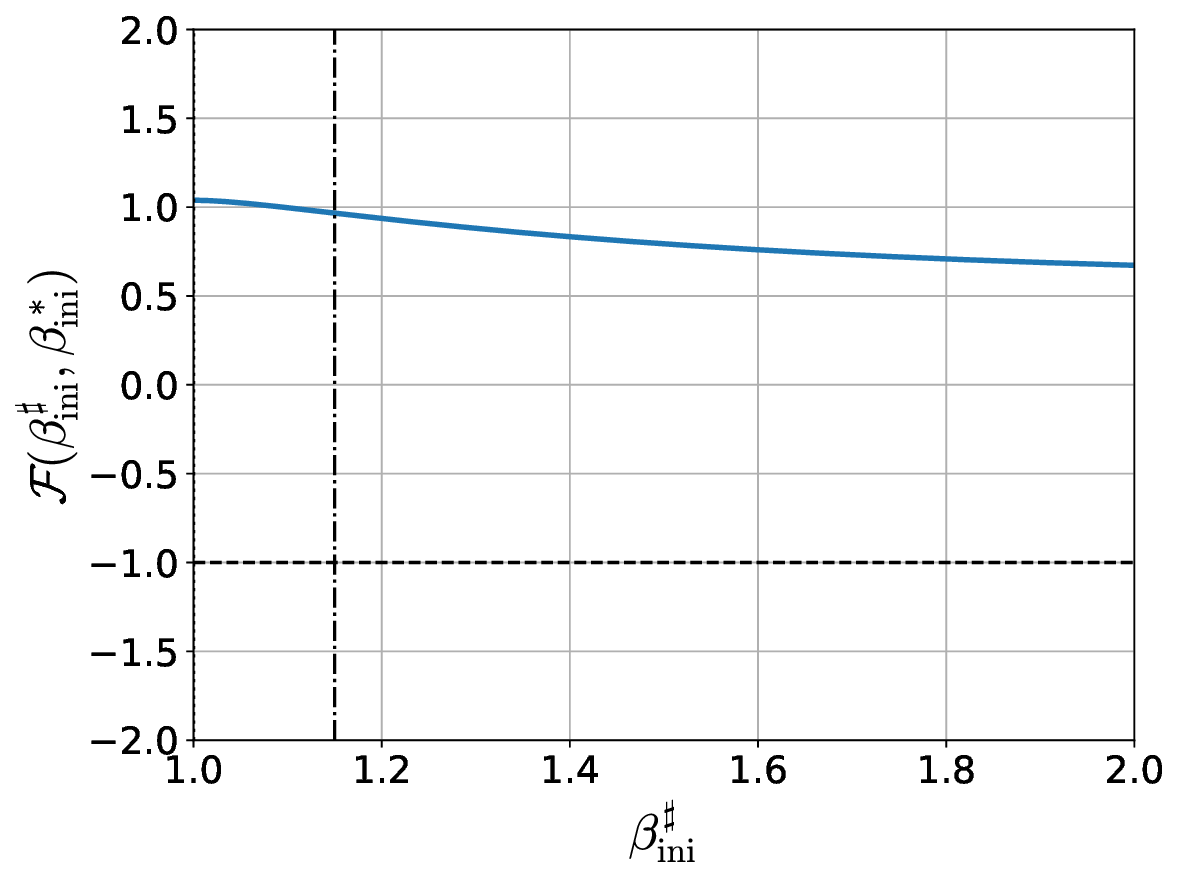}
    \caption{Plot of $\mathcal{F}(\beta_\mathrm{ini}^\sharp,\beta_\mathrm{ini}^*):=\Delta_3/\Delta_2$ against the initial temperature $\beta_\mathrm{ini}^\sharp$, where one initial temperature is fixed at $\beta^*_\mathrm{ini}=1.15$. 
    The parameter set given in Eq.~\eqref{eq:parameters_case2} is used.
    }
    \label{fig:F}
\end{figure}

For the representative parameter set, we fix the parameters as
\begin{equation}
    T=1,\quad
    k_\mathrm{in}=-k_\mathrm{mid}=k_\mathrm{out}=1,\quad
    \xi=1.0,\quad \alpha=1.84485714,
    \label{eq:parameters_case2}
\end{equation}
we plot in Fig.~\ref{fig:am_case2} the $\beta_\mathrm{ini}$
dependence of $a_m$ ($m=2,3,4,5$).
In contrast to case (i), $a_2(\beta_\mathrm{ini})$ varies monotonically with $\beta_\mathrm{ini}$ for this representative parameter set, as shown in Fig.~\ref{fig:am_case2}(a). 
Therefore, the necessary slowest-mode condition for the Mpemba effect is not satisfied.

\begin{figure}[htbp]
    \centering
    \includegraphics[width=0.5\linewidth]{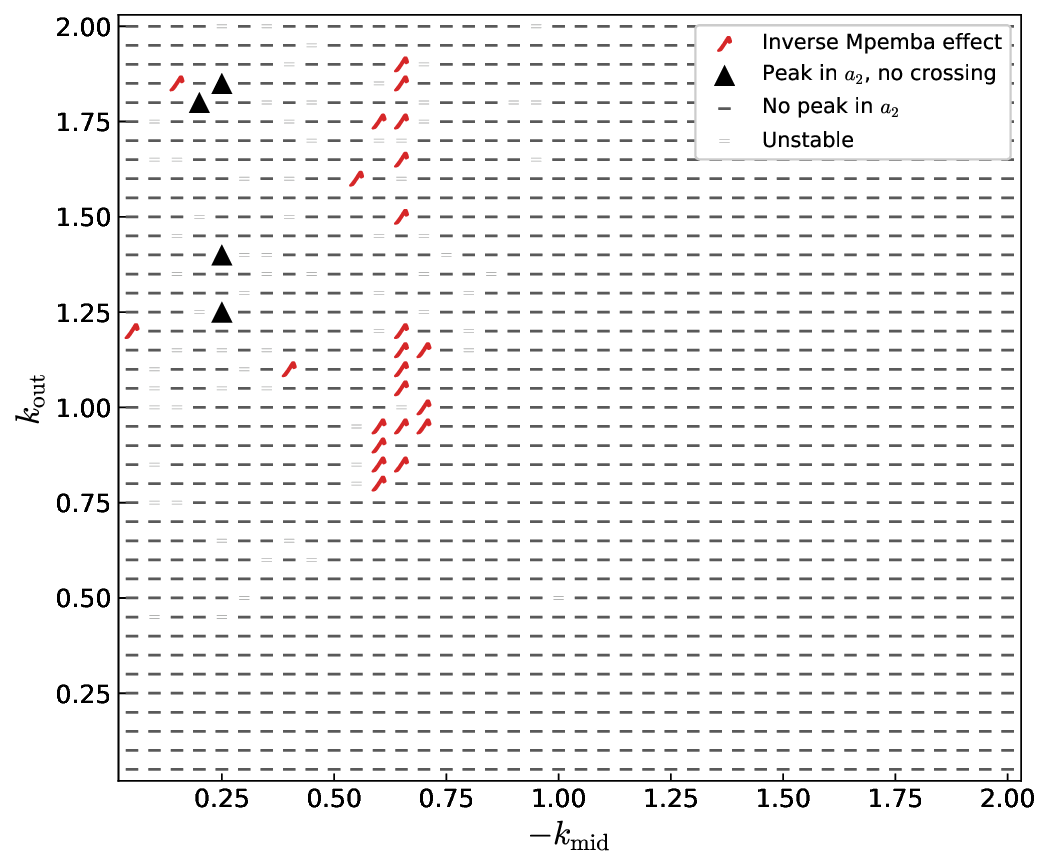}
    \caption{
    Phase diagram of the Mpemba effect in the $(-k_{\rm mid},k_{\rm out})$ plane
    for $k_{\rm in}=\xi=1$ under the marginal condition $V(\alpha)=0$.
    The red inverted check marks indicate those for which the inverse Mpemba effect is observed.
    The black triangles indicate the cases in which $a_2(\beta_{\rm ini})$ exhibits a peak
    but no crossing is found within the range of initial temperatures examined,
    whereas the gray horizontal bars indicate the cases in which
    $a_2(\beta_{\rm ini})$ has no peak.
    The light-gray double bars denote unstable cases for which
    the relaxation-mode analysis cannot be applied.
    }
    \label{fig:phase_diagram_case2}
\end{figure}

Accordingly, no Mpemba effect is expected for this representative parameter set. This is also consistent with Fig.~\ref{fig:phase_diagram_case2}, where the crossing condition is not satisfied over the examined range.
Indeed, the condition in Eq.~\eqref{A4} is not satisfied, as shown in Fig.~\ref{fig:F}, where $\mathcal{F}(\beta_\mathrm{ini}^\sharp,\beta_\mathrm{ini}^*)>-1$ throughout the range of $\beta_\mathrm{ini}^\sharp$ examined.
Thus, this example again demonstrates that the nonmonotonicity of the slowest-mode amplitude alone does not guarantee the Mpemba effect.

We next examine whether this absence of the inverse Mpemba effect is specific to the parameter set in Eq.~\eqref{eq:parameters_case2}.
Figure~\ref{fig:phase_diagram_case2} shows the phase diagram in the $(-k_{\rm mid},k_{\rm out})$ plane under the marginal condition $V(\alpha)=0$.
Here, we vary $k_{\rm mid}$ and $k_{\rm out}$ while fixing $k_{\rm in}=\xi=1$ and determine $\alpha$ for each parameter set so as to satisfy $V(\alpha)=0$.

The phase diagram reveals that the behavior on the marginal condition $V(\alpha)=0$ is more diverse than suggested by the representative parameter set above.
For most of the parameter space, $a_2(\beta_\mathrm{ini})$ has no
peak, and hence the necessary slowest-mode condition for the Mpemba
effect is absent.
There are also a few parameter sets for which $a_2$ exhibits a peak
but no crossing is found.
Remarkably, however, the inverse Mpemba effect is observed in several
localized regions of the parameter space.
These regions are relatively narrow and appear only for particular
combinations of $-k_{\rm mid}$ and $k_{\rm out}$, in contrast to the
much broader regions in which no peak in $a_2$ is found.

This result demonstrates that the marginal condition $V(\alpha)=0$
does not by itself prohibit the Mpemba effect.
Although the inverse Mpemba effect is absent for the representative
parameter set in Eq.~\eqref{eq:parameters_case2} and is relatively
rare over the parameter range examined here, it can emerge when the
curvatures of the potential are appropriately tuned.
Therefore, the equality of the two well depths is neither sufficient
to generate nor sufficient to suppress the Mpemba effect.
Rather, the detailed shape of the potential, and consequently the
competition among the relaxation modes, plays an essential role in
determining whether a crossing occurs.

Figure~\ref{fig:F} shows $\mathcal{F}(\beta_\mathrm{ini}^\sharp, \beta_\mathrm{ini}^*=1.15)$ as a function of $\beta_\mathrm{ini}^\sharp$ for the representative parameter set in Eq.~\eqref{eq:parameters_case2}.
Throughout the entire range of $\beta_\mathrm{ini}^\sharp$ examined, $\mathcal{F}$ remains larger than $-1$, confirming that the inverse Mpemba effect does not occur for this particular parameter set.

Together with the phase diagram in Fig.~\ref{fig:phase_diagram_case2} in which the inverse Mpemba effect is likely observed in some set of parameters, this result confirms the absence of the necessary slowest-mode condition for the representative parameter set, while Fig.~\ref{fig:phase_diagram_case2} shows that the inverse Mpemba effect can nevertheless emerge for other parameter choices along \(V(\alpha)=0\).
Also, there are some regions that show a peak of $a_2(\beta_\mathrm{ini})$, while neither the Mpemba nor the inverse Mpemba effect can be observed.

\section{Quantitative Criterion for Crossing of the KL Divergence}\label{App:crossing}
In this appendix, we derive a simple quantitative criterion for the
crossing of the KL divergence between two different
initial temperatures.

Consider two initial states $A$ and $B$.
Introducing 
\begin{equation}\label{Eq:B1}
    \Delta D(t) := D^A_\mathrm{KL}(t) - D^B_\mathrm{KL}(t), 
\end{equation}
with the aid of Eq.~\eqref{DKL}, we write
\begin{equation}\label{Eq:B2}
    \Delta D(t)
    =\frac{C^*}{2}
    \sum_{n\ge 1}
    \Delta_n e^{-2\lambda_n t}.
\end{equation}

A crossing occurs if there exists a time $t^\ast>0$ such that
\begin{equation}
    \Delta D(t^\ast)=0.
\end{equation}
In the minimal two-mode approximation ($n=2,3$), this gives
\begin{equation}
    \Delta_2 e^{-2\tilde{\lambda}_2 t^\ast}
    + \Delta_3 e^{-2\tilde{\lambda}_3 t^\ast}
    =0.
\end{equation}

Solving for $t^\ast$,
\begin{equation}
    t^\ast
    = \frac{1}{2(\tilde{\lambda}_3-\tilde{\lambda}_2)}
    \ln\left(\frac{-\Delta_3}{ \Delta_2}\right).
    \label{A3}
\end{equation}
A real positive solution exists if and only if Eq.~\eqref{A4} is satisfied.

Equation~(\ref{A4}) shows that a non-monotonic dependence of $a_{2}$ on the initial temperature is necessary but not sufficient for the Mpemba effect defined via KL crossing.
The amplitude contrast must be strong enough to reverse the ordering relative to faster modes.

\section{Three-mode approximation }\label{App:Higher}

In this appendix, we examine how the results change if we include higher-order terms for $\Delta D(\beta^\sharp,\beta^*):=D_\mathrm{KL}(P(\beta^\sharp, \beta;t)||P_\mathrm{eq}(\beta^\sharp))-D_\mathrm{KL}(P(\beta^*, \beta;t)||P_\mathrm{eq}(\beta))$.
Let us consider higher-order terms in the KL divergence as
\begin{align}
    D_\mathrm{KL}(P(\beta,\beta_\mathrm{ini};t)||P_\mathrm{eq}(\beta))
    &= \frac{1}{2}\sum_{n\ge 2}C^* (a_n)^2 
    e^{-2\tilde{\lambda}_n t}
    - \frac{1}{6}\sum_{\ell\ge 2}\sum_{m\ge 2}\sum_{n\ge 2} 
    D_{\ell mn}a_\ell a_m a_n
    e^{-(\tilde{\lambda}_\ell+\tilde{\lambda}_m+\tilde{\lambda}_n)t}\nonumber\\
    &\hspace{1em}
    + \frac{1}{12}\sum_{k\ge 2}\sum_{\ell\ge 2}\sum_{m\ge 2}\sum_{n\ge 2}
    E_{k\ell mn}
    a_ka_\ell a_ma_n
    e^{-(\tilde{\lambda}_k+\tilde{\lambda}_\ell+\tilde{\lambda}_m+\tilde{\lambda}_n)t}
    + O(\delta P^5),
    \label{eq:DKL_more}
\end{align}
with $D_{\ell mn}:=2\pi Z(\beta)^2 \int_0^\infty dr r e^{\beta V(r)/2}\phi_\ell(r)\phi_m(r)\phi_n(r)$ and $E_{k\ell mn}:=2\pi Z(\beta)^3\int_0^\infty dr r e^{\beta V(r)}\phi_k(r)\phi_\ell(r) \phi_m(r) \phi_n(r)$.
The first few terms of $D_{\ell mn}$ and $E_{k\ell mn}$ are given by
\begin{align}
    D_{222}&\approx 7.34748352,\ 
    D_{223}\approx 2.52984461,\
    D_{233}\approx 3.31794833,\
    D_{333}\approx -0.78230974,\\
    E_{2222}&\approx10.71271826,\ 
    E_{2223}\approx 5.09076249,\ 
    E_{2233}\approx 4.20693782,\ 
    E_{2333}\approx 2.40306438,\ 
    E_{3333}\approx 3.84968712.
\end{align}
Since $\tilde{\lambda}_n(\ge 0)$ increases with increasing $n$, we here employ the same approximation as used in Eq.~\eqref{2-mode}, keeping only $\tilde{\lambda}_2$ and $\tilde{\lambda}_3$.
Therefore, Eq.~\eqref{eq:DKL_more} becomes approximately
\begin{align}\label{Eq:C4}
    &D_\mathrm{KL}(P(\beta,\beta_\mathrm{ini};t)||P_\mathrm{eq}(\beta))=D^{(2)}(t)+D^{(3)}(t)+D^{(4)}(t)+\cdots
\end{align}
where
\begin{align}\label{eq:D^2}
D^{(2)}(t):&=\frac{1}{2} C^*\left(a_2^2 e^{-2\tilde{\lambda}_2 t}+ a_3^2 e^{-2\tilde{\lambda}_3 t}\right) ,\\
\label{eq:D^3}
D^{(3)}(t):&=
    - \frac{1}{6}
    \left(
    D_{222}a_2^3e^{-3\tilde{\lambda}_2t}
    + 3D_{223}a_2^2 a_3e^{-(2\tilde{\lambda}_2+\tilde{\lambda}_3)t}
    + 3D_{233}a_2 a_3^2e^{-(\tilde{\lambda}_2+2\tilde{\lambda}_3)t}
    + D_{333}a_3^3e^{-3\tilde{\lambda}_3t}\right) ,\\
D^{(4)}(t):&=    
    \frac{1}{12}
    \left(E_{2222}
    a_2^4e^{-4\tilde{\lambda}_2t}
    + 4E_{2223}
    a_2^3a_3e^{-(3\tilde{\lambda}_2+\tilde{\lambda}_3)t}
    + 6E_{2233}
    a_2^2a_3^2e^{-2(\tilde{\lambda}_2+\tilde{\lambda}_3)t}
    \right.\\
    &\left.
    \qquad + 4E_{2333}
    a_2a_3^3e^{-(\tilde{\lambda}_2+3\tilde{\lambda}_3)t}
    + E_{3333}
    a_3^4e^{-4\tilde{\lambda}_3t}\right).
    \label{eq:D^4}
\end{align}

Figure \ref{fig:KL-divergence} contains the dotted line as the result of Eq. \eqref{Eq:C4} and the dashed line as the numerical result with $N_\mathrm{tr}=11$, where we adopt Eq.~\eqref{2-mode} for the set of parameters.
Even if we include the higher-order contributions, the prediction of the Mpemba effect by using Eq.~\eqref{A4} is not destroyed.
Therefore, the Mpemba effect obtained in the low-order approximation persists even when higher-order contributions are taken into account.

The expansion of the KL divergence in terms of eigenmodes is useful to describe the relaxation behavior for $t>0$. 
However, for the present problem, the initial condition is an equilibrium distribution at $\beta_{\mathrm{ini}}$ [see Eq.~\eqref{P_{ini}}].
Therefore, the KL divergence at $t=0$ can be evaluated exactly from its definition Eq.~\eqref{eq:D_KL}, without relying on any mode expansion.

Equation ~\eqref{eq:D_KL} at $t=0$ can read
\begin{equation}
    D_{\mathrm{KL}}(0)
    = \int dr P_{\mathrm{eq}}(r,\beta_{\mathrm{ini}})
    \ln \frac{P_{\mathrm{eq}}(r,\beta_{\mathrm{ini}})}{P_{\mathrm{eq}}(r,\beta)}
    = (\beta - \beta_{\mathrm{ini}})\langle V \rangle_{\beta_{\mathrm{ini}}}
    + \ln \frac{Z(\beta)}{Z(\beta_{\mathrm{ini}})},
\end{equation}
where we have used Eq.~\eqref{P_{ini}}.
This is an exact expression for $\Delta D(0)$.

Thus, the difference of the KL divergences for two initial temperatures $A$ and $B$ at $t=0$ is given by
\begin{equation}
    \Delta D(0)
    = D_{\mathrm{KL}}^{A}(0) - D_{\mathrm{KL}}^{B}(0),
\end{equation}
which can be evaluated exactly from equilibrium thermodynamics.

On the other hand, for $t>0$, the deviation from equilibrium can be expressed using the spectral decomposition Eq.~\eqref{P(r,t)2}, which leads to
\begin{equation}
    D_{\mathrm{KL}}(t)
    = \frac{1}{2}\sum_{n\ge2} C_n (\hat a_n)^2 e^{-2\tilde\lambda_n t}
    + O(e^{-3\tilde\lambda_2 t}).
\end{equation}
Therefore, the time-dependent difference becomes
\begin{equation}
    \Delta D(t)
    = \Delta D(0)
    + \frac{1}{2}\sum_{n\ge2} C_n \Delta_n \left(e^{-2\tilde\lambda_n t}-1\right)
    + O(e^{-3\tilde\lambda_2 t}),
\end{equation}
where $\Delta_n := (\hat a_n^{(A)})^2 - (\hat a_n^{(B)})^2$.

A crossing requires that $\Delta D(t)$ changes sign for some $t>0$, i.e.,
\begin{align}
    \Delta D(0)\Delta D(t\to \infty)<0 .
\end{align}
Since $e^{-2\tilde\lambda_n t}-1<0$, the relaxation contributions tend to reduce the initial difference. 
Therefore, a necessary condition for crossing is that the modal contributions overcome the initial gap:
\begin{equation}
    \sum_{n\ge2} C_n \Delta_n < 0
    \quad (\text{if } \Delta D(0)>0),
\end{equation}
and similarly for the opposite case.

In the late-stage regime, keeping only the two slowest modes, we obtain
\begin{equation}
    \Delta D(t)
    \simeq \Delta D(0)
    + \frac{C^*}{2}\left[
    \Delta_2 (e^{-2\tilde\lambda_2 t}-1)
    + \Delta_3 (e^{-2\tilde\lambda_3 t}-1)
    \right].
\end{equation}
This expression provides a crossing condition when we consider many eigenmodes.

Thus, the role of higher-order modes is not to determine the initial value, but to modify how the system departs from the exact initial KL divergence. 
Their contribution can shift the crossing time quantitatively, but the qualitative mechanism of the crossing remains controlled by the competition between $\Delta D(0)$ and the slowest relaxation modes.

\section{System-size independence of the peak of $a_2(\beta_\mathrm{ini})$ in $d > 1$ dimensions}
\label{app:system_size}

In this appendix, we generalize the condition for the local extremum of the slowest mode amplitude $a_2(\beta_{\rm ini})$ to an arbitrary dimension $d > 1$ and prove that the optimal initial inverse temperature $\beta_{\rm ini}^*$ is independent of the system size (outer boundary radius) $R$, provided $R$ is sufficiently larger than the characteristic scale of the potential wells.

\subsection{Generalization to $d$ Dimensions}
In a $d$-dimensional spherically symmetric system, the volume element transforms as $d^dx = \Omega_d r^{d-1} dr$, where $\Omega_d := 2\pi^{d/2}/\Gamma(d/2)$ is the surface area of the $d$-dimensional unit sphere. The derivative of the slowest mode amplitude with respect to the initial inverse temperature, $F_2(\beta_{\rm ini}) = \partial a_2 / \partial \beta_{\rm ini}$, is expressed as:
\begin{equation}
F_2(\beta_{\rm ini}) = -\Omega_d \int_0^R dr \, r^{d-1} \phi_2(r) \left[ V(r) - \langle V \rangle_{\beta_{\rm ini}} \right] \frac{e^{-\beta_{\rm ini}V(r)}}{Z(\beta_{\rm ini})},
\end{equation}
where the partition function $Z(\beta_{\rm ini})$ and the expectation value of the potential $\langle V \rangle_{\beta_{\rm ini}}$ are given by:
\begin{equation}
Z(\beta_{\rm ini}) = \Omega_d \int_0^R dr \, r^{d-1} e^{-\beta_{\rm ini}V(r)},
\end{equation}
\begin{equation}
\langle V \rangle_{\beta_{\rm ini}} = \frac{\Omega_d}{Z(\beta_{\rm ini})} \int_0^R dr \, r^{d-1} V(r) e^{-\beta_{\rm ini}V(r)}.
\end{equation}

The peak position $\beta_{\rm ini}^*$ is defined by the vanishing condition of the covariance:
\begin{equation}
F_2(\beta_{\rm ini}^*) = 0 \implies \int_0^R dr \, r^{d-1} \phi_2(r) \left[ V(r) - \langle V \rangle_{\beta_{\rm ini}^*} \right] e^{-\beta_{\rm ini}^*V(r)} = 0.
\label{eq:peak_condition}
\end{equation}

\subsection{Asymptotic Independence of System Size $R$}
Let us consider a localized potential profile where $V(r) \to V_\infty = \text{const.}$ for $r > \alpha$, and the system is enclosed in a large \(d\)-dimensional container of radius $R \gg \alpha$.

We can split the integrals into an inner localized region $[0, \alpha]$ and an outer free region $[\alpha, R]$. In the outer region, since $V(r) = V_\infty$, the Boltzmann weight simplifies significantly. Furthermore, because the slowest relaxation mode $\phi_2(r)$ corresponds to particles transitioning out of localized potential wells into the continuum or a secondary well, its spatial profile in the flat outer region for a large system scales inversely with the system volume to maintain normalization, or exponentially decays if bounded. Crucially, the term $[V(r) - \langle V \rangle_{\beta_{\rm ini}}]$ in the outer region becomes:
\begin{equation}
V(r) - \langle V \rangle_{\beta_{\rm ini}} = V_\infty - \langle V \rangle_{\beta_{\rm ini}}.
\end{equation}

Substituting this into the peak condition Eq.~\eqref{eq:peak_condition}, we separate the integration domains:
\begin{align}
\int_0^\alpha dr \, r^{d-1} \phi_2(r) &\left[ V(r) - \langle V \rangle_{\beta_{\rm ini}^*} \right] e^{-\beta_{\rm ini}^*V(r)}
+ \left[ V_\infty - \langle V \rangle_{\beta_{\rm ini}^*} \right] e^{-\beta_{\rm ini}^* V_\infty} \int_\alpha^R dr \, r^{d-1} \phi_2(r) = 0.
\end{align}

By virtue of the orthogonality condition between the slowest mode $\phi_2(r)$ and the ground state $\phi_1(r) \propto e^{-\beta_{\rm fin}V(r)/2}$ (where $\beta_{\rm fin}$ is the final bath temperature), the total spatial integral over the entire domain vanishes when weighted appropriately. For a big or confined potential, the contribution of $\phi_2(r)$ in the outer bulk region $[\alpha, R]$ vanishes exponentially or integrates to a negligible geometric constant independent of changes in $R$. 

More directly, the partition function evaluates to:
\begin{equation}
Z(\beta_{\rm ini}) = Z_{\text{inner}}(\beta_{\rm ini}) + \Omega_d e^{-\beta_{\rm ini}V_\infty} \frac{R^d - \alpha^d}{d}.
\end{equation}
When evaluating $\langle V \rangle_{\beta_{\rm ini}^*}$, the leading-order geometric volume terms proportional to $R^d$ cancel out exactly between the numerator and denominator because $V(r) = V_\infty$ is constant across that entire outer volume. Consequently, $\langle V \rangle_{\beta_{\rm ini}^*}$ converges rapidly to a value determined strictly by the localized well structure:
\begin{equation}
\langle V \rangle_{\beta_{\rm ini}^*} \approx \frac{\int_0^\alpha dr \, r^{d-1} V(r) e^{-\beta_{\rm ini}^*V(r)}}{\int_0^\alpha dr \, r^{d-1} e^{-\beta_{\rm ini}^*V(r)}}.
\end{equation}

Since neither the local wave function $\phi_2(r)$ within the core region nor the effective energy expectation value $\langle V \rangle_{\beta_{\rm ini}^*}$ depends on the macroscopic boundary $R$, the transcendental equation determining the root $\beta_{\rm ini}^*$ is completely decoupled from $R$. Therefore, for all dimensions $d > 1$, the peak position $\beta_{\rm ini}^*$ of the slowest mode amplitude is an intrinsic property of the potential geometry and remains invariant with respect to the system size.

In this appendix, we assumed $V(r) = V_\infty$ for $r > \alpha$.
This assumption of a strictly flat potential is merely a convenient simplification and is not crucial for the system-size independence of $\beta_{\rm ini}^*$. To see this under general conditions, consider the exact vanishing condition of the covariance:
\begin{equation}
\int_0^R dr \, r^{d-1} \phi_2(r) \left[ V(r) - \langle V \rangle_{\beta_{\rm ini}^*} \right] e^{-\beta_{\rm ini}^*V(r)} = 0.
\label{eq:gen_covariance}
\end{equation}

For the root $\beta_{\rm ini}^*$ of Eq.~\eqref{eq:gen_covariance} to be invariant with respect to the macroscopic boundary $R$, it is sufficient that the integrand has localized support and vanishes rapidly in the outer region $r > \alpha$. 
Two independent physical mechanisms robustly guarantee this suppression:

\begin{enumerate}
    \item \textit{Exponential Suppresion via Boltzmann Weight:} For any physically realistic system where the potential barrier or outer region satisfies $V(r) \gg V_{\text{min}}$ (where $V_{\text{min}}$ is the depth of the wells at $r=0$ or $r=\alpha$), the Boltzmann factor $e^{-\beta_{\rm ini}^*V(r)}$ exponentially suppresses the outer integrand at low-to-intermediate initial temperatures.
    \item \textit{Localization of the Transition Mode:} The slowest relaxation mode $\phi_2(r)$ corresponds to the inter-well transport or localized bound-state transitions within the core potential geometry. Consequently, its spatial amplitude decays rapidly outside the wells (exponentially for confining boundaries, or scaling as $O(R^{-d})$ to maintain normalization in extended open spaces).
\end{enumerate}

Thus, whether $V(r)$ continues to rise (confining potential) or possesses long-range power-law tails (e.g., $V(r) \to V_\infty - C/r^n$), the contribution of the domain $[\alpha, R]$ to the integral in Eq.~\eqref{eq:gen_covariance} scales at most as an asymptotically negligible correction $O((\alpha/R)^n)$. The transcendental equation determining $\beta_{\rm ini}^*$ is therefore determined strictly by the local geometric features of the potential wells, making it highly insensitive to the specific choice or profile of $V(r)$ in the outer bulk region.

\section{Sign change of $F_2(\beta_{\rm ini})$ for the two-dimensional double-well potential}\label{app:F_2}
In this appendix, we prove that for the two-dimensional double-well potential defined in Eq.~\eqref{pot2}, $F_2(\beta_\mathrm{ini})$ introduced in Eq.~\eqref{F_2} changes sign between the limits $\beta_{\rm ini}\to 0$ and $\beta_{\rm ini}\to\infty$.
Let us rewrite $F_2(\beta_\mathrm{ini})$ as
\begin{equation}
    F_2(\beta_{\rm ini})
    =
    -2\pi
    \int_0^\infty dr r
    \phi_{2}(r)
    \left[
    V(r)-\langle V\rangle_{\beta_{\rm ini}}
    \right]
    \frac{
    e^{-(\beta_{\rm ini}-\beta/2)V(r)}
    }{Z(\beta_{\rm ini})}.
    \label{eq:F2cov}
\end{equation}

\subsection{(i) The case for $V(\alpha)<V(0)$}\label{App:D1}

Here, we assume $V(\alpha)<V(0)$ as in case (i).

\subsubsection{High-temperature limit $\beta_{\rm ini}\to 0$}

In the high-temperature limit, the initial distribution becomes broad and the measure
is dominated by the large-volume region.
Because of the Jacobian factor $r$, the dominant contribution
comes from the vicinity of the potential minima at $r=\alpha$
rather than from the barrier region near $r=0$.

For the first radial excited mode,
\begin{equation}
    \phi_{2}(r)
    \begin{cases}
    <0, & r \approx 0,\\
    >0, & r \approx \alpha.
    \end{cases}
\end{equation}
In the wells,
\begin{equation}
    V(\alpha) < \langle V\rangle_{\beta_{\rm ini}\to 0}.
\end{equation}
Hence,
\begin{equation}
    \phi_{2}(r)\,
    \big(
    V(r)-\langle V\rangle_{\beta_\mathrm{ini}}
    \big)
    <0
    \quad
    \text{in the dominant region}.
\end{equation}
Because of the overall minus sign in Eq.~(\ref{eq:F2cov}),
we conclude
\begin{equation}
F_2(0)>0.
\end{equation}

\subsubsection*{Low-temperature limit $\beta_{\rm ini}\to\infty$}

As $\beta_{\rm ini}\to\infty$, the measure concentrates near $r=\alpha$.
We therefore introduce the local coordinate
\begin{equation}
x := r-\alpha.
\end{equation}

The potential near the global minimum is expressed as
\begin{equation}\label{V(a)+}
    V(r)
    = V(\alpha)
    + \frac{k}{2}x^2,
\end{equation}
where $k=a_\mathrm{out}$.

Expand the eigenfunction:
\begin{equation}
    \phi_{2}(r)
    = c_0 + c_1 x
    + \frac{a_2}{2}x^2
    + O(x^3),
\end{equation}
where
\begin{equation}
    c_0 := \phi_{2}(\alpha).
\end{equation}
Introduce the scaled variable
\begin{equation}
    y:=\sqrt{\beta_{\rm ini} k} x ,
\end{equation}
we can write
\begin{equation}
    e^{-\beta_{\rm ini}V(r)}
    =
    e^{-\beta_{\rm ini}V(\alpha)}
    \exp\!\left(-\frac{y^2}{2}\right)
    \left[1+O(\beta_{\rm ini}^{-1/2})\right].
\end{equation}

Using Eq.~\eqref{V(a)+}, we immediately obtain
\begin{equation}\label{V-<V>}
    V(r)-\langle V\rangle_{\beta_{\rm ini}}
    = \frac{k}{2}x^2
    + O(x^3).
\end{equation}
Next, we use the expansion
\begin{align}\label{r phi_1}
    r \phi_{2}(r)
    &= (\alpha+x)
    \left(c_0+c_1 x+\frac{c_2}{2}x^2\right)
    +O(x^3)
    \\
    &= \alpha c_0
    + (\alpha c_1+c_0)x
    + \left(\frac{\alpha c_2}{2}+c_1\right)x^2
    + O(x^3).
\end{align}
Multiplying Eq.~\eqref{V-<V>} with Eq.~\eqref{r phi_1}, we obtain
\begin{align}
    r\phi_{2}(r)
    \bigl(V(r)-\langle V\rangle_{\beta_\mathrm{ini}}\bigr)
    &=
    \frac{k}{2}
    \left[
    \alpha c_0 x^2
    + (\alpha c_1+c_0)x^3
    + O(x^4)
    \right].
\end{align}

Under the Gaussian limit, we write
\begin{equation}
    \langle x^2\rangle_{\beta_\mathrm{ini}}
    =
    \frac{1}{\beta_{\rm ini} k},
    \qquad
    \langle x^3\rangle_{\beta_\mathrm{ini}}
    =
    0,
    \qquad
    \langle x^4\rangle_{\beta_\mathrm{ini}}
    =
    \frac{3}{(\beta_{\rm ini} k)^2}.
\end{equation}
Therefore, we obtain
\begin{equation}
    \int_0^\infty dr r
    \phi_{2}(r)
    \left(V(r)-\langle V\rangle_{\beta_\mathrm{ini}}\right)
    =
    \frac{\alpha c_0}{2\beta_{\rm ini}}
    +
    O(\beta_{\rm ini}^{-3/2}).
\end{equation}

We thus conclude the relation
\begin{equation}
    F_2(\beta_{\rm ini})
    =
    - \frac{\alpha\,\phi_{2}(\alpha)}{2\beta_{\rm ini}}
    + O(\beta_{\rm ini}^{-3/2})<0,
    \qquad
    \beta_{\rm ini}\to\infty.
\end{equation}

\subsubsection{Conclusion}
From the above discussion, we have established
\begin{equation}\label{F_2_two_limits}
    F_2(0)>0,
    \qquad
    \lim_{\beta_\mathrm{ini}\to \infty}F_2(\beta_\mathrm{ini})<0.
\end{equation}
By continuity, there exists at least one
$\beta_{\rm ini}^*$ such that
\begin{equation}
    F_2(\beta_{\rm ini}^*)=0.
\end{equation}
Hence, the Mpemba effect necessarily occurs
for the two-dimensional double-well potential.

\subsection{(ii) The case $V(0)<V(\alpha)$}\label{App:D2}

In this subsection, we prove the absence of the Mpemba effect for (ii) the case
\begin{equation}
    V(\alpha)>V(0)=0 .
\end{equation}
In this regime, the global minimum of the potential is located at $r=0$, while $r=\alpha$ corresponds to a metastable minimum.
Our strategy is to show $F_2(\beta_\mathrm{ini})$ introduced in Eq.~\eqref{F_2} has a definite sign for all $\beta_{\rm ini}>0$.
This excludes any extremum of $\hat a_2(\beta_{\rm ini})$, which is a necessary condition for the Mpemba effect.
The argument here can be regarded as a customized version of the original idea in Ref.~\cite{Ohga2024}, which develops a microscopic theory of the Mpemba effect.

From Eq.~\eqref{F_2} we have
\begin{equation}
    F_2(\beta_{\rm ini})
    =
    -\frac{2\pi}{Z(\beta_{\rm ini})}
    \int_0^\infty dr r
    \phi_2(r)
    e^{-\beta_{\rm ini} V(r)}
    \left[ V(r) - \langle V\rangle_{\beta_{\rm ini}} \right],
    \label{D2:raw}
\end{equation}
where
\begin{equation}
    \langle V\rangle_{\beta_{\rm ini}}
    =
    \frac{2\pi}{Z(\beta_{\rm ini})}
    \int_0^\infty dr r V(r) e^{-\beta_{\rm ini}V(r)} .
\end{equation}

Introducing the probability measure
\begin{equation}
    d\mu_{\beta_{\rm ini}}(r)
    =
    \frac{2\pi}{Z(\beta_{\rm ini})}
    r e^{-\beta_{\rm ini}V(r)} dr,
\end{equation}
Eq.~(\ref{D2:raw}) can be rewritten as
\begin{equation}
    F_2(\beta_{\rm ini})
    =
    - \mathrm{Cov}_{\beta_{\rm ini}}
    \left(\phi_2(r), V(r)\right),
    \label{D2:covform}
\end{equation}
where the covariance $\mathrm{Cov}_{\beta_{\rm ini}}\left(\phi_2(r), V(r)\right):=\phi_2(r)\left[ V(r) - \langle V\rangle_{\beta_{\rm ini}} \right]$ is taken with respect to $d\mu_{\beta_{\rm ini}}$.

Therefore, the sign of $F_2$ is opposite to the covariance between $\phi_2(r)$ and $V(r)$ under the Gibbs measure.

For $V(\alpha)>0$, the energetic ordering of the two wells is
\begin{equation}
    V(0) < V(\alpha).
\end{equation}
The first nontrivial radial eigenfunction $\phi_2(r)$ has a single node and opposite signs in the two wells.
Up to an overall normalization, we choose
\begin{equation}
    \phi_2(r)
    \begin{cases}
    <0, & r \text{ in the inner well near } 0, \\
    >0, & r \text{ in the outer well near } \alpha.
    \end{cases}
    \label{D2:signstructure}
\end{equation}

Hence, the sign of $\phi_2$ increases when moving from the lower-energy well to the higher-energy well.
In particular, for any $r_1,r_2$ belonging respectively to the inner and outer wells,
\begin{equation}
    V(r_1) < V(r_2)
    \quad\Longrightarrow\quad
    \phi_2(r_1) < \phi_2(r_2).
    \label{D2:monotone}
\end{equation}
Thus $\phi_2(r)$ is an increasing function of $V(r)$ in the sense of well-to-well ordering.

Let us decompose the integration domain into the inner-well region $\Omega_0$ and outer-well region $\Omega_\alpha$.
Because of Eq.~(\ref{D2:monotone}), $\phi_2(r)$ and $V(r)$ are positively ordered: larger values of $V$ are associated with larger values of $\phi_2$.

By Chebyshev's integral inequality for similarly ordered functions, for any probability measure (see Appendix~\ref{Chebyshev})
\begin{equation}
    \mathrm{Cov}_{\beta_{\rm ini}}
    \left(\phi_2(r), V(r)\right)
    \ge 0 .
\end{equation}

The inequality is strict because $\phi_2$ is not a constant function and the two wells have different energies.
Therefore,
\begin{equation}
    \mathrm{Cov}_{\beta_{\rm ini}}
    \left(\phi_2(r), V(r)\right)
    > 0
    \quad
    \text{for all } \beta_{\rm ini}>0 .
    \label{D2:covpositive}
\end{equation}

Importantly, the Gibbs weight
$e^{-\beta_{\rm ini}V(r)}$
changes continuously with $\beta_{\rm ini}$,
but does not alter the energetic ordering between the two wells.
Hence, the positive ordering between $\phi_2$ and $V$ is preserved for all temperatures.

Combining Eqs.~(\ref{D2:covform}) and (\ref{D2:covpositive}),
we obtain
\begin{equation}
    F_2(\beta_{\rm ini}) < 0
    \quad
    \text{for all } \beta_{\rm ini}>0 .
\end{equation}
Therefore $\hat a_2(\beta_{\rm ini})$ is strictly decreasing and has no extremum.

Since an extremum of $\hat a_2$ is a necessary condition for the crossing of the KL divergence,
the Mpemba effect cannot occur in the regime $V(\alpha)>0$.

\subsection*{Chebyshev's integral inequality for similarly ordered functions}\label{Chebyshev}

In this subsection, we state and prove Chebyshev's integral inequality
in a form appropriate for probability measures.
The proof is elementary and relies on a symmetric representation
of the covariance.

\paragraph{Setting.}

Let $(X,\mathcal{F},\mu)$ be a probability space,
i.e. $\mu(X)=1$.
Let $f,g : X \to \mathbb{R}$ be square-integrable functions.

We say that $f$ and $g$ are \emph{similarly ordered}
if for all $x,y \in X$,
\begin{equation}
    (f(x)-f(y))(g(x)-g(y)) \ge 0 .
\label{cheb:ordering}
\end{equation}
In other words, whenever $f$ increases, $g$ also increases,
and whenever $f$ decreases, $g$ also decreases.

\paragraph{Theorem (Chebyshev's integral inequality).}

If $f$ and $g$ are similarly ordered,
then
\begin{equation}
    \int_X f(x)g(x) d\mu(x)
    \ge
    \left(\int_X f(x) d\mu(x)\right)
    \left(\int_X g(x) d\mu(x)\right).
    \label{cheb:ineq}
\end{equation}
Equivalently,
\begin{equation}
    \mathrm{Cov}_\mu(f,g) \ge 0 .
\end{equation}
If the inequality in \eqref{cheb:ordering}
is strict on a set of positive $\mu\times\mu$ measure,
then the covariance is strictly positive.

\paragraph{Proof.}
Recall that the covariance can be written symmetrically as
\begin{equation}
    \mathrm{Cov}_\mu(f,g)
    = \frac{1}{2}
    \int_X \int_X
    (f(x)-f(y))(g(x)-g(y))
    d\mu(x) d\mu(y).
    \label{cheb:symmetric}
\end{equation}

To verify \eqref{cheb:symmetric}, expand the integrand:
\begin{align}
    &(f(x)-f(y))(g(x)-g(y)) \nonumber\\
    &= f(x)g(x) + f(y)g(y) - f(x)g(y) - f(y)g(x).
\end{align}
Integrating over $X\times X$ and using $\mu(X)=1$,
we obtain
\begin{align}
    \int_X\int_X f(x)g(x) d\mu(x)d\mu(y)
    &= \int_X f(x)g(x) d\mu(x),\\
    \int_X\int_X f(x)g(y) d\mu(x)d\mu(y)
    &= \left(\int_X f d\mu\right)
       \left(\int_X g d\mu\right),
\end{align}
and similarly for the remaining terms.
Collecting all contributions yields
\begin{equation}
    \int_X\int_X
    (f(x)-f(y))(g(x)-g(y))
    d\mu(x)d\mu(y)
    =
    2\mathrm{Cov}_\mu(f,g),
\end{equation}
which proves \eqref{cheb:symmetric}.

Now assume that $f$ and $g$ are similarly ordered,
i.e. \eqref{cheb:ordering} holds for all $x,y$.
Then the integrand in \eqref{cheb:symmetric}
is nonnegative everywhere.
Therefore,
\begin{equation}
    \mathrm{Cov}_\mu(f,g) \ge 0.
\end{equation}

If the inequality in \eqref{cheb:ordering}
is strict on a set of positive $\mu\times\mu$ measure,
then the double integral is strictly positive,
and hence
\begin{equation}
    \mathrm{Cov}_\mu(f,g) > 0.
\end{equation}

This completes the proof.
\hfill $\square$



\end{document}